\documentclass[journal=jpcafh,manuscript=article]{achemso}
\SectionNumbersOn
\usepackage{amsmath}
\usepackage{braket}
\usepackage{graphicx}
\usepackage{subcaption}
\usepackage{booktabs}
\usepackage[american]{babel}
\usepackage{multirow}

\usepackage{listings}
\usepackage{xcolor}

\definecolor{codegreen}{rgb}{0,0.6,0}
\definecolor{codegray}{rgb}{0.5,0.5,0.5}
\definecolor{codepurple}{rgb}{0.58,0,0.82}
\definecolor{backcolour}{rgb}{0.95,0.95,0.92}

\lstdefinestyle{mystyle}{
    backgroundcolor=\color{backcolour},   
    commentstyle=\color{codegreen},
    keywordstyle=\color{magenta},
    numberstyle=\tiny\color{codegray},
    stringstyle=\color{codepurple},
    basicstyle=\ttfamily\footnotesize,
    breakatwhitespace=false,         
    breaklines=true,                 
    captionpos=b,                    
    keepspaces=true,                 
    numbers=left,                    
    numbersep=5pt,                  
    showspaces=false,                
    showstringspaces=false,
    showtabs=false,                  
    tabsize=2
}

\usepackage[colorlinks,linkcolor=black,urlcolor=blue, citecolor=black]{hyperref} 

\title[QCMaquis 4.0]{
  QCMaquis 4.0: Multi-Purpose Electronic, Vibrational, and Vibronic Structure and Dynamics Calculations with the Density Matrix Renormalization Group
}

\author{Kalman Szenes\footnote{ORCID: 0009-0006-3642-0674}}
\affiliation{ETH Zurich, Department of Chemistry and Applied Biosciences, Vladimir-Prelog-Weg 2, 8093 Zurich, Switzerland}
\author{Nina Glaser\footnote{ORCID: 0000-0002-7477-8099}}
\affiliation{ETH Zurich, Department of Chemistry and Applied Biosciences, Vladimir-Prelog-Weg 2, 8093 Zurich, Switzerland}
\author{Mihael Erakovic\footnote{ORCID: 0000-0002-1879-510X}}
\affiliation{ETH Zurich, Department of Chemistry and Applied Biosciences, Vladimir-Prelog-Weg 2, 8093 Zurich, Switzerland}
\author{Valentin Barandun\footnote{ORCID: 0009-0003-1236-4119}}
\affiliation{ETH Zurich, Department of Chemistry and Applied Biosciences, Vladimir-Prelog-Weg 2, 8093 Zurich, Switzerland}
\author{Maximilian M{\"o}rchen\footnote{ORCID: 0000-0002-7467-5719}}
\affiliation{ETH Zurich, Department of Chemistry and Applied Biosciences, Vladimir-Prelog-Weg 2, 8093 Zurich, Switzerland}
\author{Robin Feldmann\footnote{ORCID: 0000-0003-2038-3072}}
\affiliation{ETH Zurich, Department of Chemistry and Applied Biosciences, Vladimir-Prelog-Weg 2, 8093 Zurich, Switzerland}
\author{Stefano Battaglia\footnote{ORCID:  0000-0002-5082-2681}}
\affiliation{Department of Chemistry, University of Zurich, Winterthurerstrasse 190, Zurich, 8057, Switzerland}
\author{Alberto Baiardi\footnote{ORCID:  0000-0001-9112-8664}}
\affiliation{ETH Zurich, Department of Chemistry and Applied Biosciences, Vladimir-Prelog-Weg 2, 8093 Zurich, Switzerland}
\author{Markus Reiher\footnote{ORCID: 0000-0002-9508-1565}}
\affiliation{ETH Zurich, Department of Chemistry and Applied Biosciences, Vladimir-Prelog-Weg 2, 8093 Zurich, Switzerland}
\email{mreiher@ethz.ch}

\date{4/30/2025}

\begin{document}

\begin{abstract}
    QCMaquis is a quantum chemistry software package for general molecular structure calculations in a matrix product state/matrix product operator formalism of the density matrix renormalization group (DMRG).
    It supports a wide range of features for electronic structure, multi-component (pre-Born--Oppenheimer), anharmonic vibrational structure, and vibronic calculations.
    In addition to the ground and excited state solvers, QCMaquis allows for time propagation of matrix product states based on the tangent-space formulation of time-dependent DMRG.
    The latest developments include transcorrelated electronic structure calculations, very recent vibrational and vibronic models, and a convenient Python wrapper, facilitating the interface with external libraries.
    This paper reviews all the new features of QCMaquis and demonstrates them with new results. 
\end{abstract}

\section{Introduction}
Among the broad range of quantum chemical methods established in the past decades, tensor network algorithms have emerged as a particularly promising group of high-accuracy approaches for efficiently treating strongly correlated quantum systems~\cite{ma22_dmrgbook,marti10_review-dmrg,Verstraete2023May,White1998Jul,orus19,Murg2010Nov,xiang2023density}.
These methods leverage suitable tensor factorization schemes to efficiently represent and manipulate complex quantum many-body wavefunctions that are otherwise beyond the reach of traditional numerical techniques.
Various tensor network algorithms have been developed across diverse scientific disciplines. 
Examples include tree tensor networks~\cite{Murg2010Nov,Nakatani2013Apr,Larsson2019Nov}, with promising applications in chemistry, the projected entangled pair states 
(PEPS)~\cite{verstraete04} and the multiscale entanglement renormalization ansatz (MERA)~\cite{Vidal08_MERA}, commonly utilized for lattice systems in condensed matter physics, or the multi-configurational time-dependent Hartree (MCTDH)~\cite{meyer90_mctdh} approach for molecular quantum dynamics simulations.
Despite differences in target applications and algorithmic implementations, these approaches all rely on systematic tensor factorization schemes to efficiently capture the Hamiltonian's essential features.

In quantum chemistry, the most successful tensor-based algorithm is the density matrix renormalization group (DMRG)~\cite{white92_dmrg,white93_dmrgbasis,Ostlund1995Nov}, which relies on a one-dimensional tensor network topology referred to as matrix product state (MPS)
or tensor train (TT)~\cite{oseledets12_als}.
So-called area laws~\cite{hastings07_arealaw} guarantee that the ground state wavefunction of one-dimensional short-ranged gapped Hamiltonians can be efficiently encoded in a TT factorization.
This allows DMRG to deliver high accuracy while taming the inherently exponential computational 
cost of solving the Schr\"odinger equation.
Initially devised to study spin chains in solid-state physics, the algorithm was subsequently applied to Hamiltonians that are neither inherently (pseudo-)one-dimensional nor limited to short-range interactions, such as the ones found in electronic structure theory~\cite{shuai96_original,fano1998density,shuai98_PPP,white99_qcdmrg,chan02,legeza03_dmrg-lif-optimalordering,marti08_dmrg-tmc}.
For such systems, area laws no longer apply.
Nonetheless, the DMRG algorithm has been shown to be a highly efficient method for the deterministic variational optimization of the electronic wavefunction~\cite{marti08_dmrg-tmc}.
Today, it is widely regarded as a benchmark method for large-scale electronic structure calculations of strongly correlated systems such as transition metal complexes and clusters~\cite{legeza08_review,chan08_dmrgreview,chan09_dmrgreview,marti10_review-dmrg,schollwoeck11_dmrgrev,chan11_dmrgreview,wouters13_dmrgreview,kurashige14_dmrgreview,olivaresamaya15_dmrginpractice,szalay15_review,yanai15_dmrgrev,knecht16_chimia,baiardi20_dmrgreview,freitag20_bookchapter,ma22_dmrgbook}.

The linear one-dimensional TT factorization employed in DMRG is among the simplest tensor network topologies.
Despite this apparent simplicity, MPSs have proven to be remarkably versatile, successfully tackling a broad range of applications beyond their originally intended purpose and often rival state-of-the-art specialized networks.
In fact, the simplicity of the MPS structure avoids many of the challenges associated with more complex tensor network topologies.
For instance, higher-dimensional tensor networks, such as the two-dimensional PEPS networks or the complete-graph and related tensor networks \cite{Marti2010Oct,Kovyrshin2016Nov,Kovyrshin2017Dec}, can better capture the correlation structure of certain systems but become computationally demanding for increasing lattice dimension.
Furthermore, DMRG benefits from well-established and extensively tested schemes based on quantum information metrics~\cite{Barcza2011Jan} to resolve the arbitrariness of the mapping of the quantum system (e.g., described in terms of orbitals) to the linear tensor network lattice.
This remains an active field of research in more complex tensor network topologies, such as tree tensor networks~\cite{Murg2015Mar,Larsson2019Nov,mendivetapia23_modecomb} due to the increased number of possible site permutations.

DMRG offers several advantages over competing stochastic approximate solvers, such as Full Configuration Interaction Quantum Monte Carlo (FCIQMC)~\cite{Booth2009Aug}.
One key benefit is that, in its standard optimization scheme, DMRG is variational, guaranteeing that the energy is always an upper bound to the exact energy.
Furthermore, the cost and accuracy of a DMRG calculation are primarily governed by a single controllable parameter, namely the bond dimension, which can be systematically increased to refine the accuracy and converge the results toward the exact solution.
Additional parameters that affect the accuracy of the method have been summarized in Refs.~\citenum{keller14_dmrgparams,olivaresamaya15_dmrginpractice}.
Its robust optimization scheme allows for the application of DMRG as an 'off-the-shelf' method for studying target systems without needing in-depth knowledge of the algorithm's inner workings.
This accessibility has contributed to the widespread adoption of DMRG.

In recent years, the DMRG algorithm has been extended beyond its original role as a ground-state solver for time-independent problems.
The convenient tensor network formulation of DMRG~\cite{schollwoeck11_dmrgrev} facilitates the direct application of standard linear algebra techniques to wavefunctions expressed as an MPS.
This has led to the development of several DMRG-based algorithms, which, instead of computing the ground state solution, target excited states~\cite{rakhuba16_vdmrg,baiardi19_highenergy_vdmrg,baiardi22_dmrg-feast,dorando2007Davidson,hu2015maxO,liao2023transcorrelated_excited_states,yu2017highly_excited,khemani2016highly_excited,devakul2017highly_excited,rakhuba2016MPS_excited,mcculloch2007ortho}.
Furthermore, various time-dependent formulations of DMRG have emerged~\cite{Greene2017Sep,frahm19_td-dmrg_ultrafast,Holtz2012Mar,lubich14_timeintegrationtt,haegeman16_mpo-tddmrg}, enabling the study of quantum dynamic processes.
Remarkably, these time-dependent approaches often demonstrate comparable computational cost and accuracy~\cite{Mainali2021May,ren22_td-dmrg,Larsson2024Jul} to established quantum dynamics methods, such as MCTDH. 

Alongside these algorithmic developments, DMRG has been extended to tackle quantum many-body Hamiltonians beyond traditional spin models or ab initio electronic structure.
As a result, DMRG has been developed toward anharmonic vibrational structure~\cite{baiardi17_vdmrg,baiardi19_highenergy_vdmrg,glaser23_nmode_vdmrg,Larsson2019Nov,Larsson2024Jul,ren21_td-dmrg-nmode}, 
electronic~\cite{baiardi21_electrondynamics,Wahyutama2024Nov} and vibronic~\cite{baiardi19_td-dmrg,sheng2024td, larsson20242500, ye2021constructing, li2020finite,ren18_tddmrg-temperature,yao18_td-dmrg,wang2021internal_conversion} quantum dynamics, rotational Hamiltonians~\cite{iouchtchenko2018ground}, 
treatment of finite-temperature effects~\cite{feiguin2005finite}, open quantum systems~\cite{nuomin2024open_system_dmrg,prior2010open_quantum_dmrg,LeDe2024open_quantum_dmrg}, and multi-component systems, such as in nuclear-electronic~\cite{muolo20_neap-dmrg,feldmann22_prebo_dmrg} and polaritonic chemistry~\cite{Matousek2024Jul}.

Since its inception more than 30 years ago, DMRG has inspired the development of several mature open-source implementations, often specializing in a specific field of application.
One of the most widely used tensor network packages is ITensor~\cite{ITensor}, which has strong roots in condensed matter physics and is primarily designed for lattice systems.
Other notable and widely used packages include quimb~\cite{gray2018quimb} and TeNPy~\cite{hauschild2018efficient}.
In electronic structure theory, a few DMRG packages are available.
Notable examples include the various versions of the Block program (Block, StackBlock~\cite{sanshar2024Dec}, and, recently, Block2~\cite{Zhai2023Dec}) from the Chan group and the ChemMPS2~\cite{Wouters2014Jun} package, which, however, is no longer under active development.
Both software stacks interface with various quantum chemistry packages, such as OpenMolcas~\cite{Fdez.Galvan2019Nov,Aquilante2020Jun,LiManni2023Oct} and PySCF~\cite{Sun2018Jan,Sun2020Jul}.
In addition, the groups of Legeza and Veis have developed massively parallelized implementations of the DMRG algorithms for electronic structure calculations on GPUs~\cite{Menczer2024Oct,Menczer2024Oct2} and CPUs~\cite{Brabec2021Mar}, but these programs remain closed source and are not publicly available.
For applications involving vibrational and vibronic systems, the Renormalizer~\cite{shuaigroup2025Feb} package from the Shuai and the recently unveiled Kylin-V~\cite{Xu2024Aug} program from the Ma group are some of the few open-source options.

We have developed the open-source QCMaquis package in the last decade, which started as an electronic structure program~\cite{keller15_mpo,Keller2016Apr} and has soon been extended toward vibrational structure~\cite{baiardi17_vdmrg,baiardi19_highenergy_vdmrg}.
QCMaquis offers a comprehensive set of algorithms designed to tackle quantum chemical problems across domains, providing several unique features that set it apart from other DMRG programs.

Here, we present version 4.0 of the program, which collects various previous developments, combined with significant enhancements to the program's capabilities.
Notable extensions to electronic structure calculations include support for explicitly correlated calculations through the transcorrelated method~\cite{baiardiExplicitlyCorrelatedElectronic2022,szenesStrikingRightBalance2024} and an interface to the complete-active-space second-order perturbation theory (CASPT2)~\cite{anderssonSecondorderPerturbationTheory1990,Andersson1992Jan} method in OpenMolcas, 
relying solely on up to three-body reduced density matrices (RDMs).
A new vibrational model leverages the $n$-mode quantized Hamiltonian~\cite{glaser23_nmode_vdmrg}, which introduces generic modal basis sets for the optimization of vibrational energies, requiring integrals of the $n$-mode potential with respect to the chosen basis set.
For common basis sets --- such as harmonic oscillator eigenfunctions, one-body potential eigenfunctions, and vibrational self-consistent field (VSCF)
modals --- these integrals may be computed with our Colibri software~\cite{Colibri1.0.0}.
In addition, modal correlation analyses based on quantum information metrics, including the generation of modal entanglement diagrams, can be calculated for vibrational systems~\cite{glaser24_vib-ent}.
New vibronic Hamiltonians have also been integrated into the program; among them is the Frenkel excitonic Hamiltonian, which accommodates the description of modes connecting neighboring monomers.
Finally, a new Python interface for interacting with QCMaquis has been developed, which facilitates the program integration into complex workflows.
This interface can be used as a drop-in replacement for active space solvers in the PySCF quantum chemistry package.

This work is structured as follows:
Section~\ref{sec:theory} reviews fundamental theoretical aspects of the DMRG algorithm.
Afterwards, various features of the QCMaquis program are presented, each accompanied by a new DMRG demonstration calculation on an illustrative example.
First, the available Hamiltonians---spin-lattice, electronic, vibrational, and vibronic---are outlined in Section~\ref{sec:hamiltonians}. Then, excited state solvers are discussed in Section~\ref{sec:excited}.
Subsequently, Section~\ref{sec:time-dependent} presents the tangent-space formulation of time-dependent DMRG for both real- and imaginary-time evolution.
Sections~\ref{sec:dynamic-corr} and \ref{sec:interfaces} focus on electronic structure calculations and discuss dynamic correlation methods and interfaces to external quantum chemistry programs, respectively.
Tools for wavefunction analyses, such as $n$-particle and $n$-orbital RDMs, quantum information metrics, and autocorrelation functions are outlined in Section~\ref{sec:measurements}, followed by a brief discussion of technical aspects of the program in Section~\ref{sec:technical}.
Finally, the paper describes the latest version of the AutoCAS program in Section~\ref{sec:autocas}, which automatically determines active orbital spaces and can serve as an automated driver of QCMaquis. This paper closes with concluding remarks and an outlook on future developments.

\section{Theory}\label{sec:theory}
The QCMaquis program leverages the matrix product state/matrix product operator (MPS/MPO) formulation of the DMRG algorithm~\cite{schollwoeck11_dmrgrev}.
In this formalism, the wavefunction coefficient tensor is decomposed into a tensor network comprised of a linear chain of $L$ rank-3 site tensors.
Each of these tensors is assigned to a local degree of freedom (DoF) of the system.
Through the occupation number vector (ONV) representation in second quantization, these DoFs refer to orbitals in electronic structure theory or modals in vibrational structure theory.
Moreover, composite schemes have been developed in which individual tensors are assigned different particle types, as, for instance, in vibronic systems, composed of vibrational and electronic DoFs, or in pre-Born--Oppenheimer nuclear-electronic schemes, which treat both electrons and nuclei on the same footing.

In general, the wavefunction ansatz is then given by an MPS,
\begin{equation}
  \ket{\Psi} = \sum_{\boldsymbol{\sigma}} \sum_{\boldsymbol{\alpha}} M_{\alpha_{1}}^{\sigma_{1}} M_{\alpha_{1},\alpha_{2}}^{\sigma_{2}} \ldots M_{\alpha_{L-1}}^{\sigma_{L}} \ket{\sigma_{1}\sigma_{2}\ldots\sigma_{L}},
\label{eq:mps}
\end{equation}
composed of a linear chain of rank-3 tensors (except for the first and last tensor, which are matrices), where the index $\sigma_i$ corresponds to the occupancy of DoF $i$, and $\alpha_i$ introduces (non-physical) auxiliary indices along which neighboring tensors are contracted.
If the extent of the $\alpha_i$ indices, referred to as the \textit{bond dimension}, is allowed to grow exponentially along the length of the chain, this factorization can exactly represent any arbitrary quantum state from the system's entire Hilbert space (albeit at the cost of becoming completely impractical).
In practice, the bond dimension is set to a fixed maximal value, leading to an algorithm that scales polynomially with system size.

In the MPS/MPO formalism, operators are also represented as one-dimensional tensor networks (that is, MPOs),
\begin{equation}
  \hat{O} = \sum_{\boldsymbol{\sigma}\boldsymbol{\sigma'}} \sum_{\boldsymbol{\alpha}} W_{\alpha_{1}}^{\sigma_{1}\sigma_{1}'} W_{\alpha_{1},\alpha_{2}}^{\sigma_{2}\sigma_{2}'} \ldots W_{\alpha_{L-1}}^{\sigma_{L}\sigma_{L}'} \ket{\sigma_{1}\sigma_{2}\ldots\sigma_{L}}\bra{\sigma_{1}'\sigma_{2}'\ldots\sigma_{L}'}.
\end{equation}
Unlike the MPS, the MPO usually corresponds to an exact factorization of the underlying operator.
Various approaches have been proposed to compress the operator further~\cite{chanMatrixProductOperators2016,hubig17_genericmpo}.
These techniques rely on truncating the singular value decomposition of the individual MPO site tensors.
In this process, the sparsity of the MPO representation is usually compromised~\cite{hubig17_genericmpo}.
To preserve the sparsity of the MPO, QCMaquis adopts an exact MPO factorization for all operators.

The MPO factorization of an operator is inherently non-unique. Although deriving a naive MPO representation is straightforward, constructing a compact representation---characterized by a small bond dimension---remains a challenge, especially for operators containing long-range interactions and $n$-body coupling terms.
To address this challenge, several algorithms have been proposed for constructing compact MPOs of arbitrary operators~\cite{chanMatrixProductOperators2016,hubig17_genericmpo,Ren2020Aug}, including the method developed by our group~\cite{keller15_mpo}, which has been implemented in the first version of QCMaquis in 2015.
Our procedure begins by constructing an initial, naive MPO representation of the operator from its symbolic second-quantized form.
It then systematically identifies and consolidates recurring substrings of creation and annihilation operators across the operator’s terms, reducing redundancy in the MPO tensors and achieving a more compact representation.
For a comprehensive explanation of our algorithm, the reader is directed to our original work in Ref.~\citenum{keller15_mpo}.

One crucial aspect of an efficient DMRG implementation is the exploitation of global symmetries imposed by the Hamiltonian.
These symmetries can come in the form of particle number or spin conservation and have the effect of decomposing the tensors in the tensor networks into distinct sectors characterized by their respective quantum numbers~\cite{McCulloch2002Mar,vidal11_dmrg-u1symm}, resulting in a block-sparse structure of the MPS and MPO tensors.
While this block sparsity significantly reduces computational and memory costs, it also complicates the implementation of efficient tensor contraction schemes, especially for non-Abelian symmetries like the SU(2) symmetry imposed by spin conservation~\cite{Menczer2024Oct}.
As a result, modern DMRG programs must be carefully engineered to fully leverage these sparsity patterns and achieve optimal performance.

The success of DMRG stems not only from its compact representation of wavefunctions and operators but also from its efficient and robust optimization scheme. 
This approach provides a mechanism for optimizing each MPS site tensor individually while keeping the others fixed, thereby reducing the effective size of the problem.
The optimization progresses through a series of "sweeps" across the lattice, solving each site's resulting local optimization problem until convergence.
Typically, this local problem consists of an eigenvalue equation or, in the case of certain excited state solvers, a linear system of equations for the MPS site tensor under consideration.
In practice, the "two-site" variant of DMRG is usually employed.
In this method, two neighboring sites are contracted along their shared index to form a "supersite", which is then optimized, followed by its decomposition back into its original two site tensors using the singular value decomposition.
Although this two-site approach is computationally more demanding than the single-site variant, it provides several practical advantages.
The larger local problem reduces the likelihood of getting trapped in local minima, making the optimization more robust. 
Furthermore, the discarded singular values during the supersite decomposition can be used to estimate the error of the method, helping to determine the required bond dimension for a given accuracy and enabling extrapolation to the infinite bond dimension~\cite{Legeza1996Jun,chan02}.
Due to these advantages, unless stated otherwise, the two-site variant of DMRG was employed for all numerical simulations contained in this paper.

\section{Available Hamiltonians}\label{sec:hamiltonians}

QCMaquis implements facilities for treating several classes of Hamiltonians occurring in model spin systems, electronic structure theory, vibrational and vibronic problems, and multi-component nuclear-electronic systems.
These Hamiltonians are detailed in the sections below.

\subsection{Spin Lattice Hamiltonians}

\subsubsection{Fermi-Hubbard Model}

QCMaquis provides facilities for simulating the two-dimensional lattice Fermi--Hubbard model Hamiltonian in both the real-space representation
\begin{align}
  \mathcal{H}_{\text{FH}}^{(r)} =t \sum_{\braket{\mathbf{i},\mathbf{j}}}\sum_{\sigma} a^{\dagger}_{\mathbf{i},\sigma}a_{\mathbf{j},\sigma} + U \sum_{\mathbf{i}}n_{\alpha\mathbf{i}} n_{\beta,\mathbf{i}},
\end{align}
where $\mathbf{i} = (i_{x}, i_{y})$ and $\braket{ \mathbf{i}, \mathbf{j}}$ corresponds to nearest neighbor sites, and in momentum space representation
\begin{align}
  \mathcal{H}_{\text{FH}}^{(m)} = -t \sum_{\mathbf{k},\sigma} \epsilon_{\mathbf{k}} n_{\mathbf{k},\sigma} + U \sum_{\mathbf{p},\mathbf{q},\mathbf{k},\sigma} c^{\dagger}_{\mathbf{p}-\mathbf{k},\sigma} c^{\dagger}_{\mathbf{q}+\mathbf{k},\overline{\sigma}} c_{\mathbf{q},\overline{\sigma}} c_{\mathbf{p}, \sigma},
\end{align}  
where the creation operators $a_{\mathbf{i},\sigma}$ have been transformed to the momentum-space operators using a unitary transformation
\begin{align}
    c_{\mathbf{k},\sigma} = \frac{1}{\sqrt{WL}} \sum_{\mathbf{i}} e^{i \mathbf{k} \cdot \mathbf{i}} a_{\mathbf{i},\sigma} .
\end{align}

In addition to standard calculations on the model, QCMaquis can perform explicitly correlated calculations based on the transcorrelated~\cite{Francis1969Apr,Handy1969Oct} formalism.
In this method, the wavefunction is described by an ansatz given by the product of a correlator $e^{\tau}$ and a determinantal expansion, which, in our case, is provided by an MPS
\begin{align}
  \ket{\psi_{\text{tc}}} = e^{\tau} \ket{\psi_{\text{MPS}}}.
\end{align}
The correlator is incorporated into the Hamiltonian through the use of a similarity transformation, yielding a non-Hermitian operator $\mathcal{\tilde{H}}_{\text{tcFH}} = e^{-\tau}\mathcal{H}_{\text{FH}}e^{\tau}$.
In the case of the Fermi--Hubbard model, the correlator is chosen to be the Gutzwiller correlator~\cite{dobrautz2019compact,baiardi20_tcdmrg}, which improves the description of electron correlation by penalizing configurations containing doubly occupied sites.

\subsection{Electronic Structure Theory}

\subsubsection{Conventional electronic Hamiltonian}

QCMaquis can perform electronic structure calculations using the Born--Oppenheimer electronic Hamiltonian given in atomic units by
\begin{align}
  \mathcal{H}_\text{el} = - \frac{1}{2}\sum_i \nabla_i^2 - \sum_{i, I} \frac{Z_I}{r_{iI}} + \sum_{j < i} \frac{1}{r_{ij}} + \sum_{I,J} \frac{Z_I Z_J}{r_{IJ}}
\label{eq:1stQuantElectronicHamiltonian}
\end{align}
with $i,j$ and $I,J$ corresponding to electronic and nuclear indices, respectively, and $Z_I$ describing the charge of nucleus $I$.
QCMaquis can be used to target a specific irreducible representation of the molecular point group, an eigenstate of the $S_z$ operator, or an $S^2$ spin configuration using a spin-adapted approach~\cite{Sharma2012Mar,Wouters2014Jun,Keller2016Apr}.

\subsubsection{Transcorrelated Hamiltonian}
QCMaquis also supports performing explicitly correlated calculations using the transcorrelated method~\cite{Francis1969Apr,Handy1969Oct}, which
aims to reduce the basis set incompleteness error by incorporating prior knowledge of the wavefunction into a modified Hamiltonian $\mathcal{H}_{\text{tc}}$.
This is achieved by performing a similarity transformation of the electronic Hamiltonian $\mathcal{H}_{\text{el}}$ using a function $F$, known as the correlator,
\begin{align}\label{eq:tc-sim-trans}
  \mathcal{H}_{\text{tc}} = F^{-1} \mathcal{H}_{el} F.
\end{align}
This correlator is typically parametrized by an exponential, whose arguments explicitly depend on the inter-particle distances
\begin{align}
  F = e^\tau 
\end{align}
with
\begin{align}
\tau = \sum_{i < j} f(r_{ij}),
\end{align}
where $i$ and $j$ correspond to electronic indices.
Similar to R12/F12 methods~\cite{kongExplicitlyCorrelatedR122012}, the correlator is primarily chosen such that the analytic cusp conditions~\cite{katoEigenfunctionsManyParticle1957,packCuspConditionsMolecular1966} of the exact wavefunctions at the point of coalescence of two charged particles are satisfied.
The similarity transformation in Eq.~(\ref{eq:tc-sim-trans}) can be expanded using the Baker--Campbell--Hausdorff formula, which naturally truncates after the first nested commutator
\begin{align}
  \nonumber
  \mathcal{H}_\text{tc} = e^{-\tau}\mathcal{H}_\text{el}e^{\tau} &= \mathcal{H}_\text{el} + [\mathcal{H}_\text{el}, \tau] + \frac{1}{2}[[\mathcal{H}_\text{el}, \tau], \tau]\\
  &= \mathcal{H}_\text{el} - \sum_i \frac{1}{2}\nabla_i\tau + \nabla_i \tau \cdot \nabla_i + \frac{1}{2} (\nabla_i \tau)^2. \label{eq:tc-bch}
\end{align}
This transformation results in a non-Hermitian operator that includes up to three-body interactions due to the third and fourth terms in Eq.~(\ref{eq:tc-bch}), respectively.
In second-quantized form $\mathcal{H_\text{tc}}$ is expressed as
\begin{align}
\mathcal{H}_\text{tc} =& \sum_{\mu \nu} \sum_{s}^{\left\{ \alpha,\beta \right\}}
    h_{\mu \nu} a_{\mu s}^\dagger a_{\nu s}
  + \frac{1}{2} \sum_{\mu \nu \lambda \sigma} \sum_{s s'}^{\left\{ \alpha,\beta \right\}}
    \widetilde{( \mu \nu | \lambda \sigma )} a_{\mu s}^\dagger a_{\lambda s'}^+ a_{\sigma s'} a_{\nu s} \\
&- \frac{1}{6} \sum_{\mu \nu \lambda \sigma \kappa \tau} \sum_{s s' s''}^{\left\{ \alpha,\beta \right\}}
   K_{\mu\nu,\lambda\sigma,\kappa\tau} 
   a_{\mu s}^\dagger a_{\lambda s'}^\dagger a_{\kappa s''}^\dagger a_{\tau s''} a_{\sigma s'} a_{\nu s},
\end{align}
where $\widetilde{( \mu \nu | \lambda \sigma )}$ are modified two-body integrals and $K_{\mu\nu,\lambda\sigma,\kappa\tau}$ are additional three-body integrals arising from the similarity transformation.
The expressions for these integrals can be found in Ref.~\citenum{baiardiExplicitlyCorrelatedElectronic2022}.
The steep increase in computational cost due to the three-body term may be tamed by normal-ordering the Hamiltonian with respect to a reference state~\cite{liaoEfficientAccurateInitio2021,christlmaierXTCEfficientTreatment2023}, usually the Hartree--Fock solution.
The normal ordering then allows contributions from the three-body operator to be incorporated into lower-body terms so that the remaining---usually small---pure three-body fluctuation potential can be neglected.
The resulting Hamiltonian is expressed as
\begin{align}
  \mathcal{H}_{\text{tc-NO}} & = E_\text{ref} + \{\mathcal{H}_{1B}\} + \{\mathcal{H}_{2B}\},
\end{align}
where the bracketed operators denote the normal-ordered one- and two-body operators with respect to the reference state $\ket{\Psi_{\text{ref}}}$ and $E_\text{ref} = \braket{\Psi_\text{ref}| \mathcal{H}_\text{tc} | \Psi_\text{ref}}$ corresponds to the energy of the reference state.
The validity of this approximation has been confirmed on a wide array of systems~\cite{liaoEfficientAccurateInitio2021,schraivogelTranscorrelatedCoupledCluster2021,christlmaierXTCEfficientTreatment2023,szenesStrikingRightBalance2024} and since $\mathcal{H}_{\text{tc-NO}}$ has only up to two-body interactions, it retains the same computational scaling as $\mathcal{H}_\text{el}$.
The transcorrelated calculations can be performed with the conservation of the number spin $\alpha$ and $\beta$ electrons.

\subsubsection{Four-component Relativistic Hamiltonians}

While spin-averaged scalar-relativistic effects on electronic structures can be efficiently addressed at the level of one-electron integrals~\cite{Moritz2005Nov,Peng2012Jan},
QCMaquis also includes a relativistic four-component formulation based on time-independent Hamiltonians in the absence of external magnetic fields~\cite{Battaglia2018a}.
The relativistic DMRG model is based on the Dirac-Coulomb Hamiltonian, with the optional inclusion of the Breit interaction for the approximate description of magnetic and retardation effects in the interaction of two electrons~\cite{reiher2014relativistic}.
The four-component spinor formulation encompasses both relativistic kinematic effects and spin-orbit coupling, which
enables a variational treatment of relativistic effects directly in the optimization of the MPS wavefunction.
We note that for scalar-relativistic (one-component) methods, spin-orbit coupling may also be taken into account in a subsequent step by matrix product state state-interaction~\cite{Knecht2016Dec}.
Approximate two-component Hamiltonians, such as 
exact two-component~\cite{Liu10072010,Saue2011Dec} 
and zero-order regular approximation~\cite{Chang1986,Lenthe1993,Lenthe1994} 
are also supported.

The implementation is based on a Kramers pair basis $\{\varphi_l, \bar{\varphi_l}\}$, built from pairs of fermion spinor functions $\{ \varphi_l \}$ and $\{ \bar{\varphi_l} \}$, respectively.
In the absence of external magnetic fields, these are doubly degenerate according to Kramers' theorem and related by the time-reversal operator.
However, in practice, each spinor is considered independently in QCMaquis, thereby each lattice site is 2-dimensional with the spinor being either unoccupied or occupied.

Based on a quaternion symmetry scheme, the full range of Abelian double group symmetries is also supported.
In addition to the $\text{D}_{2h}^*$ group and subgroup thereof, QCMaquis also supports the finite groups $\text{C}_{16h}^*$ and $\text{C}_{32v}^*$, with and without inversion symmetry, to be used as approximations of the full groups $\text{D}_{\infty h}^*$ and $\text{C}_{\infty v}^*$ for linear molecules. 
As the number of particles is preserved, relativistic calculations can be run with combined U(1) and double group symmetry.

\subsection{Nuclear-Electronic Pre-Born--Oppenheimer Hamiltonian}

A unique feature of QCMaquis, not available in any other DMRG software, is the implementation of the pre-Born--Oppenheimer model.
In this model, nuclei and electrons are treated on an equal footing, allowing for either all nuclei to be treated fully quantum mechanically, or selected nuclei to be represented by classical point charges. 
Potential future applications of this model are high-precision studies of processes involving coupled quantum mechanical motion of electrons and nuclei, such as proton-coupled electron transfer reactions and anharmonic spectroscopy. 
In addition to the orbitals from conventional electronic structure theory, including strongly correlated nuclear orbitals significantly expands the system's active space, rapidly exceeding the limits of exact diagonalization methods and highlighting the necessity of approximate active space solvers like DMRG.
Currently, the lack of reliable subsequent multicomponent approaches for accounting for dynamic correlation within large active spaces limits the accuracy of pre-Born--Oppenheimer methods.
Despite this limitation, the DMRG implementation in QCMaquis provides a crucial foundation for advancing these techniques in the future.

If all nuclei are treated quantum mechanically, the pre-Born--Oppenheimer Hamiltonian corresponds to the full molecular Hamiltonian, expressed as the sum of the electronic Hamiltonian from Eq.~(\ref{eq:1stQuantElectronicHamiltonian}) and the kinetic energy of the nuclei 
\begin{equation}
    \mathcal{H}_{\mathrm{preBO}} = \mathcal{H}_\mathrm{el} - \sum_A \frac{1}{2m_A} \nabla_A^2,
\end{equation}
where $m_A$ is the mass of nucleus $A$.
Representing this Hamiltonian in second quantization is a non-trivial task, as it requires accounting for the bosonic or fermionic nature of a given nucleus and incorporating all possible spin quantum numbers.
Specifically, this would include all integers (including zero) for bosons and all positive half-integers for fermions.
The current implementation in QCMaquis is limited to spin-0 bosons and spin-$\frac{1}{2}$ fermions.
The Hamiltonian for a system of spin-$\frac{1}{2}$ fermions is given by
\begin{equation}
 \mathcal{H}_\text{preBO} = \sum_I^{N_t} \sum_{ij}^{L_I}  \sum_{s=\uparrow,\downarrow}
  h_{Iij}~ \hat{a}^\dagger_{Is, i} \hat{a}_{I s,j}
  + \frac{1}{2} \sum_{IJ}^{N_t} \sum_{ik}^{L_I}
  \sum_{jl}^{L_J} \sum_{ss^\prime=\uparrow,\downarrow}
  V_{Iik,Jjl} ~ 
  \hat{a}^\dagger_{Is, i} \hat{a}^\dagger_{Js^\prime,j}   
  \hat{a}_{J s^\prime, l } \hat{a}_{I s, k} ~.
  \label{eq:SQPreBO}
\end{equation}
Here, $h_{Iij}$ and $V_{Iik,Jjl}$ are the one- and two-body integrals calculated over spatial molecular orbitals.
$N_t$ is the number of distinguishable particle types with corresponding capital indices, while
$L_I$ represents the number of orbitals for particle type $I$.
Lowercase indices correspond to orbital indices, and the spin variable $s$ can be spin-up or down for spin-$\frac{1}{2}$ fermions.
The implementation of the model accounts for the distinct symmetries of different particle types and ensures the commutation of operators associated with different particle types.
This is achieved by adapting the Jordan--Wigner transformation as detailed in Ref.~\citenum{muolo20_neap-dmrg}.

\subsection{Anharmonic Vibrational Systems}
In addition to electronic and pre-Born--Oppenheimer Hamiltonians, QCMaquis can also be utilized to perform anharmonic vibrational calculations.
It supports two of the most widely used vibrational models: Watson-type~\cite{hirata2014normal} and $n$-mode~\cite{christiansen2004second} Hamiltonians.

\subsubsection{Watson Hamiltonian}
Watson-type Hamiltonians can be applied to potential energy surfaces (PESs) that are represented by general sum-over-product expressions of position and momentum operators.
\begin{align}
  \mathcal{H}_{\text{Watson}} &=  \frac{1}{2} \sum_{i=1}^M \omega_i^2 \left( \hat{P}_i^2 + \hat{Q}_i^2 \right)
  + \sum_{i=1}^M k_i\hat{Q}_i + \sum_{i=1}^M l_i\hat{P}_i \notag
  + \frac{1}{2}\sum_{ij}f_{ij}\hat{Q}_i\hat{P}_j 
  + \frac{1}{2}\sum_{ij}g_{ij}\hat{P}_i\hat{Q}_j \notag \\
  &\quad + \frac{1}{6} \sum_{ijk} k_{ijk} \hat{Q}_i \hat{Q}_j \hat{Q}_k
  +\; \dots \, , \label{eq:watson}
\end{align}
where $M$ is the number of vibrational modes, and $\hat{Q}_i$ and $\hat{P}_i$ are the position and momentum operators associated with the $i$-th mass-weighted normal mode.
Our QCMaquis code can treat Hamiltonian terms of $\hat{Q}_i$ and $\hat{P}_i$ of arbitrary order, including mixed expressions for rovibrational couplings.
The operators $\hat{Q}_i$ and $P_i$ are expressed in terms of bosonic creation $\hat{b}_i^\dagger$ and annihilation $\hat{b}_i$ operators as $\hat{Q}_i = \frac{1}{\sqrt{2}} \left( \hat{b}_i^\dagger + \hat{b}_i \right)$ and $\hat{P}_i = \frac{i}{\sqrt{2}} \left( \hat{b}_i^\dagger - \hat{b}_i \right)$.

\begin{figure}[htb]
    \centering
    \includegraphics[width=0.35\linewidth]{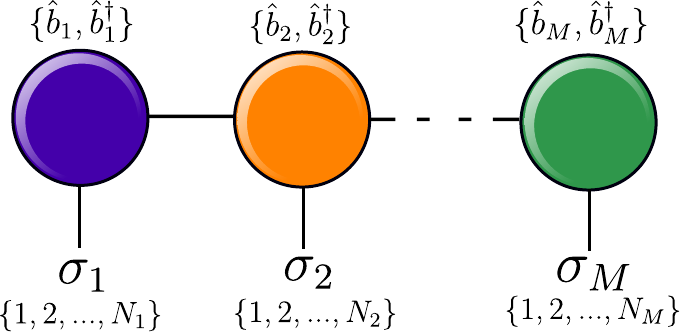}
    \caption{
    Tensor diagram of the canonical vibrational lattice.
    Each site corresponds to a vibrational mode, indicated by the different colors of the tensor sites, with its corresponding bosonic harmonic oscillator creation and annihilation operators, $\hat{b}_i, \hat{b}^{\dagger}_i$, positioned above.
    The local basis $i$ is determined by the first $N_i$ vibrational states, given in curly brackets.
    }
    \label{fig:canonical-lattice}
\end{figure}

These second-quantization operators define the mapping of the Hamiltonian, and correspondingly also the vibrational wavefunction, onto the DMRG lattice.
Specifically, for the Watson-type Hamiltonian, each lattice site corresponds to a vibrational normal mode $i$, and the wavefunction is consequently expanded in terms of harmonic oscillator eigenfunctions, where a maximum number of $N_i$ harmonic oscillator eigenfunctions is chosen as the local basis of each site $i$.
This mapping of vibrational DoF to DMRG lattice sites is referred to as \textit{canonical} and is depicted in Fig.~\ref{fig:canonical-lattice}.
All possible combinations of occupied vibrational basis functions are physically allowed since the site occupations are independent of each other.
This lattice, therefore, does not exhibit any particle number conservation symmetry.

\subsubsection{$n$-Mode Hamiltonian} \label{subsec:nmode_hamil}

\begin{figure}[htb]
    \centering
    \includegraphics[width=0.7\linewidth]{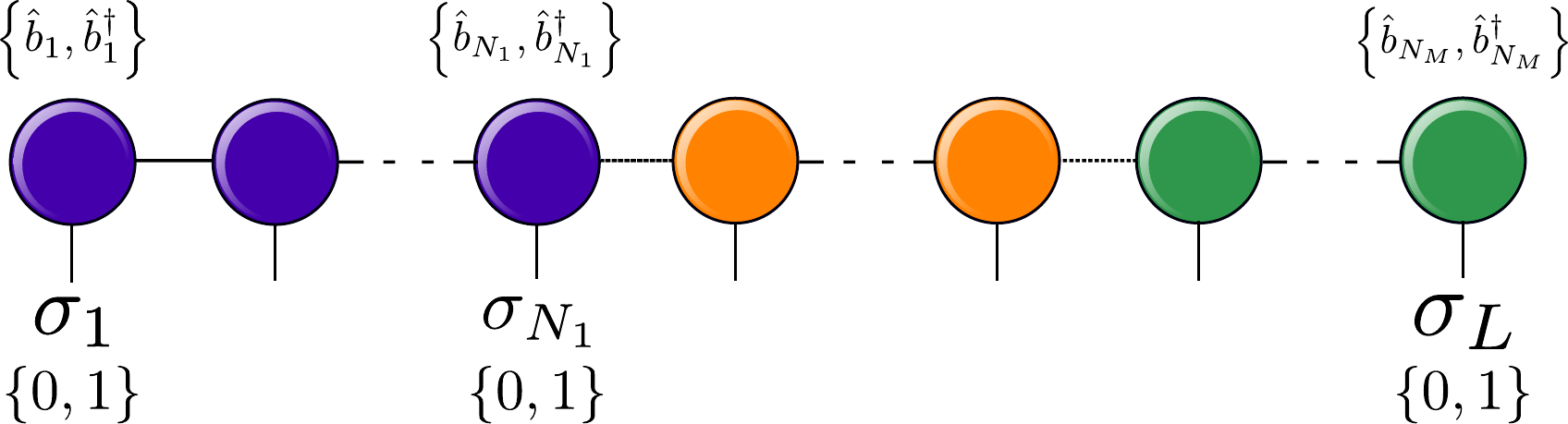}
    \caption{
    Tensor diagram of the $n$-mode vibrational lattice, where each site corresponds to a modal basis function.
    The modals are color-coded according to their associated vibrational mode.
    The local dimension of each site is two, reflecting the occupancy of the modal, with the constraint that any single configuration only possesses a single occupied modal per normal mode.
    Shown above the tensor sites are their respective bosonic creation and annihilation operators. 
    }
    \label{fig:longLattice}
\end{figure}

Alternatively, the vibrational DMRG algorithm can also be applied to anharmonic systems described by the more flexible $n$-mode expansion of the PES with the following Hamiltonian, neglecting rotational coupling terms,
\begin{equation}\label{eq:hamilton_nmode_first_quant}
  \mathcal{H}_{\text{nmode}} = \sum\limits_{i=1}^M \mathcal{T} (Q_i) + \sum\limits_{i=1}^M \mathcal{V}_1^{[i]}(Q_i) + \ldots + \sum\limits_{i<j<\ldots}^M \mathcal{V}_n^{[i,j,\ldots]}(Q_i, Q_j, \ldots) \, ,
\end{equation}
where $\mathcal{T}$ is the kinetic energy operator.
The $n$-body potential terms $\mathcal{V}_n$ depend at most on $n$ of the $M$ normal modes, which do not need to be in product form or adhere to any specific functional format.
The $n$-mode Hamiltonian can be expressed in second quantization using a generic anharmonic modal basis set as
\begin{equation}\label{eq:hamilton_nmode}
  \mathcal{H}_{\text{nmode}} =  \sum_{i=1}^M \sum_{k_i,h_i=1}^{N_i} H_{k_i, h_i}^{[i]}
  \hat{b}_{k_i}^\dagger \hat{b}_{h_i}
  + \sum_{i=1}^M\sum_{i<j}^{M} \sum_{k_i,h_i=1}^{N_i} \; \sum_{k_j,h_j=1}^{N_j} H_{k_i k_j, h_i h_j}^{[i,j]}
  \hat{b}_{k_i}^\dagger \hat{b}_{k_j}^\dagger \hat{b}_{h_i} \hat{b}_{h_j} + \ldots \, ,
\end{equation}
where the one-body integrals $H_{k_i, h_i}^{[i]}$ contain both the kinetic and potential one-mode contributions of mode $i$, the two-body integrals $H_{k_i k_j, h_i h_j}^{[i,j]}$ contain two-mode potential contributions, and analogous expressions for higher-order terms follow.
The creation $\hat{b}_{k_i}^\dagger$ and annihilation $\hat{b}_{k_i}$ operators are defined with respect to the $k_i$-th basis function $\phi_i^{k_i}$ associated with the $i$-th mode.
The $n$-mode Hamiltonian thus maps each vibrational mode to several modal basis functions $\phi_i^{k_i}$, each corresponding to a site on the DMRG lattice.
This lattice is illustrated in Fig.~\ref{fig:longLattice}.

\subsection{Electronic and Vibrational Coupling}
\label{sec:vibronic}

QCMaquis can be applied to simulate vibronic processes, where the interplay between electronic and vibrational DoFs is essential for capturing the underlying physical phenomena.
It provides two classes of vibronic Hamiltonians.
The first one models the dynamics of Frenkel excitons along molecular aggregates~\cite{kouppel1984multimode,fradkin1983phase,von1995hubbard}.
The second describes a general class of systems~\cite{raab1999molecular} containing arbitrary electronic and vibrational DoF that are coupled together.
Typical applications of this model include the computation of vibrationally resolved electronic spectra.

\begin{figure}[htb]
    \centering
    \includegraphics[width=0.7\linewidth]{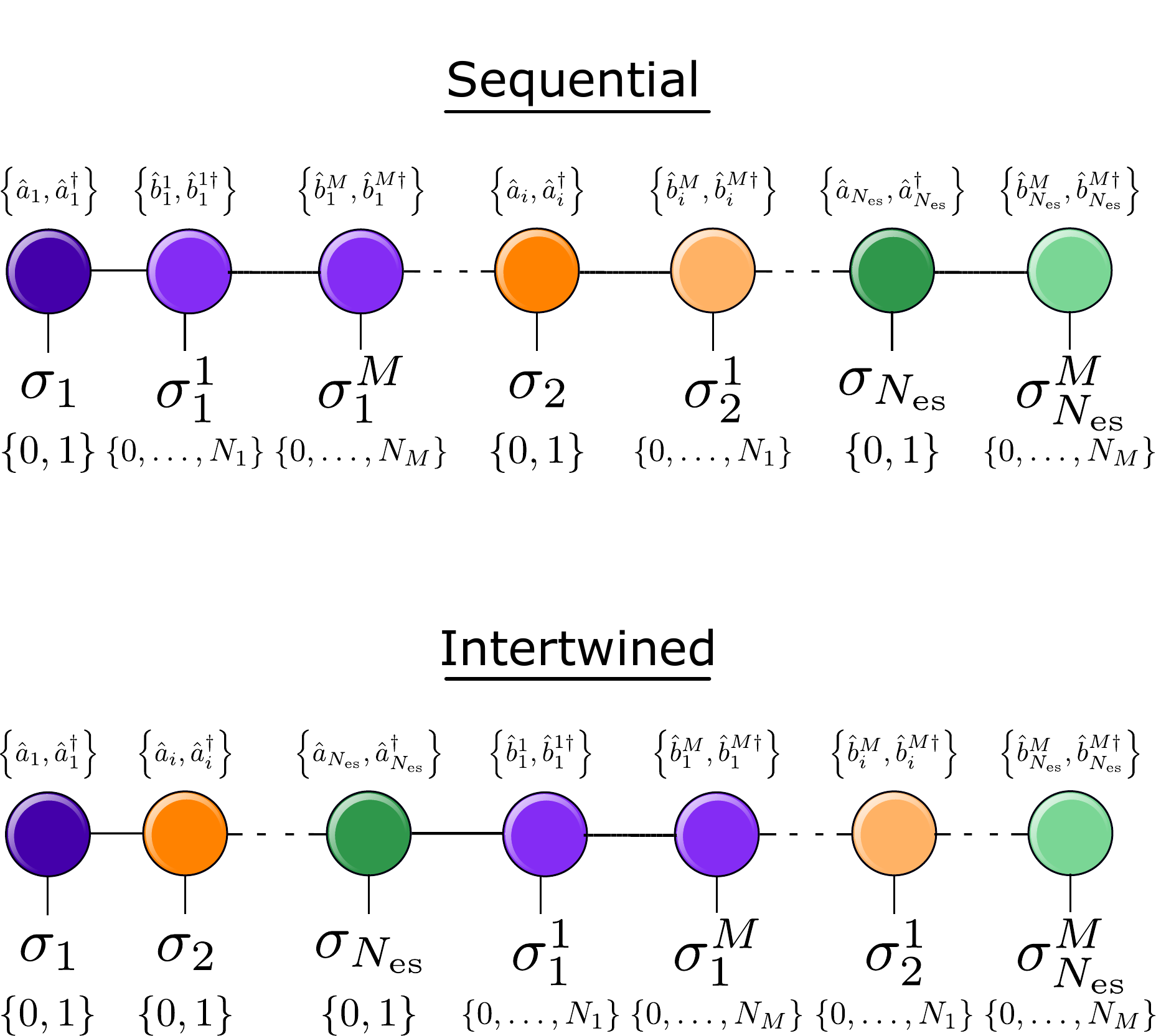}
    \caption{
    Tensor diagrams for the two lattices supported by QCMaquis for vibronic calculations.
    Sites are color-coded according to their electronic states, with corresponding vibrational modes shown in a lighter shade.
    Above each site tensor, the operators ($\hat{a}$ for fermionic and $\hat{b}$ for bosonic) corresponding to the degrees of freedom (DoF) are indicated, while the possible occupancy of these DoFs is listed below.
    The sequential lattice first includes sites related to the electronic, followed by the vibrational DoFs.
    In contrast, the intertwined lattice ordering groups all DoFs associated with the same electronic state.
    }
    \label{fig:vibronic-lattice}
\end{figure}

The two supported MPS lattice orderings for vibronic calculations are depicted in Fig.~\ref{fig:vibronic-lattice}.
The sequential lattice lists all electronic DoFs before the vibrational DoFs, while the intertwined lattice orders DoFs based on their corresponding electronic state and, thus, interleaves electronic and vibrational DoFs.

\subsubsection{Frenkel Excitonic Systems}
To efficiently describe molecular aggregates composed of multiple identical chromophores, each described by the same electronic states and vibrational DoFs, the Frenkel excitonic model Hamiltonian can be used. 
In QCMaquis, the Frenkel excitonic model is comprised of two electronic states, which typically include the ground state, for each chromophore.
The vibrational DoFs are described in terms of harmonic oscillator PESs and the Watson Hamiltonian $\mathcal{H}_\text{Watson}$ for the ground and the excited electronic state.
This model assumes nearest-neighbor interactions with a unique scalar coupling term $J_\text{coupl}$ between neighboring monomers.
Given $N_\text{mon}$ monomers with $N_\text{vib}$ vibrational modes each, the excitonic Hamiltonian can be written as
\begin{align}\label{eq:hamilton_exciton}
  \nonumber
  \mathcal{H}_\text{exc} = 
  \frac{1}{2} &\sum_{i=1}^{N_{\text{mon}}} \sum_{j=1}^{N_{\text{vib}}} \omega_{j}^2 \left( \hat{P}_{i,j}^2 + \hat{Q}_{i,j}^2 \right) \ket{\mathbf{0}} \bra{\mathbf{0}} +
  \sum_{i=1}^{N_\text{mon}}
  \mathcal{H}_\text{Watson} (Q_{i,1}, \dots, Q_{i,N_{\text{vib}}}) \vert\Psi_{S_i}\rangle \langle\Psi_{S_i}\vert
  + \\
  + &J_\text{coupl}\sum_{i=1}^{N_\text{mon}-1}
   \left( \vert\Psi_{S_i}\rangle \langle\Psi_{S_{i+1}}\vert + \vert\Psi_{S_{i+1}}\rangle \langle\Psi_{S_i}\vert \right),
\end{align}
where the indices $i$ and $j$ run over the monomers and vibrational mode, respectively, with $\hat{Q}_{i, j}$ and $\hat{P}_{i, j}$ corresponding to the $j$-th vibrational coordinate and corresponding momentum of monomer $i$.
The frequency $\omega_j$ corresponds to the frequency of the vibrational mode $j$ and is assumed identical for all monomers.
The first double sum in Eq.~(\ref{eq:hamilton_exciton}) corresponds to the harmonic oscillator Hamiltonian of the ground electronic state $\ket{\mathbf{0}}$ of the aggregate.
The second term describes, using $\mathcal{H}_\text{Watson}$, the vibrational Hamiltonian of the aggregate's excited electronic states $\ket{\Psi_{S_i}}$, which correspond to a configuration with all monomers except for monomer $i$ in the electronic ground state.
The last term accounts for the nearest-neighbor coupling between electronic states.

\subsubsection{Generic Vibronic Processes}

For an arbitrary vibronic system containing $N_\text{es}$ electronic states, the Hamiltonian takes the form
\begin{equation}
  \mathcal{H}_\text{vibronic} =
  \begin{bmatrix}
    \mathcal{H}_1(\mathbf{Q})
    & \mathcal{V}_{12}(\mathbf{Q})
    & \cdots
    & \mathcal{V}_{1 N_\text{es}}(\mathbf{Q}) \\
    \mathcal{V}_{21}(\mathbf{Q})
    & \mathcal{H}_2(\mathbf{Q})
    &
    & \\
    \vdots
    &
    & \ddots
    & \\
    \mathcal{V}_{N_\text{es} 1}(\mathbf{Q})
    &
    &
    & \mathcal{H}_{N_\text{es}}(\mathbf{Q}) \\
  \end{bmatrix},
  \label{eq:VibronicHamiltonian}
\end{equation}
where $\mathbf{Q}=(Q_1,\ldots,Q_{M})$ denotes the $M$ vibrational DoFs.
The diagonal terms $\mathcal{H}_{m}(\mathbf{Q})$ denote the vibrational Hamiltonian associated with the $m$-th electronic state, and is expressed as
\begin{equation}
  \mathcal{H}_{m}(Q_1,\ldots,Q_{M}) = E_m^{(\text{eq})}
  + \frac{1}{2} \sum_{j=1}^{M} \omega_j^{(m),2} \left( Q_j^2  + P_j^2 \right)
  + \sum_{j=1}^{M} g_{j}^{(m)} Q_j \, ,
  \label{eq:DiagonalVibronic}
\end{equation}
where $E_m^{(\text{eq})}$ stands for the electronic energy of the $m$-th electronic state at the ground state equilibrium molecular structure, $Q_j$ and $P_j$ refer, respectively, to the $j$-th dimensionless normal coordinate and corresponding conjugate momentum, with corresponding harmonic frequency $\omega_j$ and linear shift coefficient $g_{j}^{(m)}$.
The off-diagonal terms $\mathcal{V}_{mn}(\mathbf{Q})$ represent the non-adiabatic couplings between the $m$-th and the $n$-th electronic state, and can be expressed as
\begin{equation}
  \mathcal{V}_{mn}(\mathbf{Q}) =
  \sum_{k=1}^{M} g_k^{(m,n)} Q_k
  + \sum_{k,l=1}^{M} h_{k,l}^{(m,n)} Q_k Q_l \, ,
  \label{eq:OffDiagonalVibronic}
\end{equation}
with $g_k^{(m,n)}$ and $h_{k,l}^{(m,n)}$ corresponding to the first- and second-order non-adiabatic coupling coefficients, respectively.

\section{Available Excited-State Algorithms}\label{sec:excited}

QCMaquis provides several excited state solvers that are applicable to any of the models outlined above.
These algorithms are detailed in the following sections.

\subsection{Sequential Low-Lying Excited States with DMRG[ORTHO]}

The most straightforward extension of the standard DMRG algorithm to target excited states is the DMRG[ORTHO] variant.
In this approach, excited states are calculated sequentially by enforcing orthogonality of the current MPS to all previously calculated MPSs, which correspond to lower-lying states~\cite{mcculloch2007ortho,keller15_mpo,baiardi17_vdmrg}.
This is achieved through a constrained optimization procedure that minimizes the energy of the current MPS in the subspace orthogonal to the lower-lying states. Such procedure formally corresponds to the ground state optimization of the modified Hamiltonian 
\begin{equation}
    \tilde{\mathcal{H}} = \mathcal{P}_{\{\ket{\phi_n}\}^{\perp}}\mathcal{H}\mathcal{P}_{\{\ket{\phi_n}\}^{\perp}},
\label{eq:projected_hamiltonian}
\end{equation}
where $\mathcal{P}_{\{\ket{\phi_n}\}^{\perp}}$ corresponds to the projection operator onto the orthogonal complement of the subspace spanned by the set of lower-lying MPSs $\{\ket{\phi_n}\}$.
Although DMRG[ORTHO] is practical and efficient for computing low-energy excited states, the method suffers from some significant drawbacks when applied to higher-lying excited states
as it requires computing all lower-lying states before reaching the target state, leading to increased computational cost and error accumulation as higher excited states are targeted.

\subsection{Arbitrary Excited States using the Inverse Power Iteration with DMRG[IPI]}

The DMRG[IPI] algorithm leverages the inverse power iteration (IPI) approach\cite{saadNumericalMethodsLarge2011,rakhuba16_vdmrg} to compute excited states by repeatedly applying the shifted-and-inverted Hamiltonian on the guess wavefunction
\begin{equation}
  \vert \Psi_k \rangle = \left( \mathcal{H} - \omega \right)^{-1} \vert \Psi_{k-1} \rangle \, ,
  \label{eq:ipi1}
\end{equation}
at each iteration $k$.
The expensive explicit inversion of the modified Hamiltonian is avoided by reformulating the operation as $\Gamma_{\omega} \vert \Psi_k \rangle = \vert \Psi_{k-1} \rangle$ where $\Gamma_{\omega} = \left( \mathcal{H} - \omega \right)$.
In practice, the wavefunction is optimized by minimizing the functional
\begin{equation}
  \tilde{O}_{\omega} \left[ \Psi_{k} \right] = \langle \Psi_k \vert \Gamma_{\omega} \vert \Psi_k \rangle  - 2 \langle \Psi_{k-1} \vert \Psi_k \rangle \, .
  \label{eq:residualMinimum}
\end{equation}
in a sweeping procedure, yielding the optimal MPS of a given bond dimension at iteration $k$~\cite{baiardi22_dmrg-feast}.
Assuming that the initial guess MPS $\vert \Psi_0 \rangle$ has non-vanishing overlap with the targeted eigenstate, the wavefunction $\vert \Psi_k \rangle$ converges with increasing $k$ towards the eigenstate of the Hamiltonian whose energy lies closest to the shift parameter $\omega$.
Each DMRG[IPI] calculation requires an energy shift $\omega$ parameter provided by the user.

\subsection{Solving Entire Energy Intervals using DMRG[FEAST]}

To efficiently compute excited states within densely populated regions of the eigenspectrum, DMRG[FEAST] is a particularly powerful approach, as an entire energy interval can be computed at once~\cite{baiardi22_dmrg-feast}.
This method is based on the FEAST algorithm~\cite{polizzi09}, which simultaneously computes all eigenfunctions within a specified energy range $I_E = [ E_{\text{min}}, E_{\text{max}} ]$ through an iterative subspace diagonalization procedure.
An initial set of $M$ linearly independent guess states $\{\Psi_\text{guess}^{(1)}, \ldots, \Psi_\text{guess}^{(M)}\}$ is projected to the subspace spanned by the eigenfunctions of the Hamiltonian contained in the interval $I_E$.
Using Cauchy's integral theorem, this projector $ \mathcal{P}_{M} $ can be expressed as a complex contour integral
\begin{equation}
  \mathcal{P}_{M} = \sum_{i=1}^{M}  \vert \Psi^{(i)} \rangle \langle \Psi^{(i)} \vert
  = \frac{1}{2\pi \mathrm{i}} \oint_{\mathcal{C}} (z - \mathcal{H})^{-1} dz \, ,
  \label{eq:FEAST_Projector}
\end{equation}
which, in practice, is approximated with an $N_\text{p}$-point numerical quadrature:
\begin{equation}
  \mathcal{P}_{M} \ket{\Psi^{(i)}_{\text{guess}}}
  \approx \frac{1}{2\pi \mathrm{i} } \sum_{k=1}^{N_\text{p}} w_k (z_k  - \mathcal{H})^{-1} \ket{\Psi^{(i)}_{\text{guess}}}
  = \frac{1}{2\pi \mathrm{i} } \sum_{k=1}^{N_\text{p}} w_k \ket{\Psi^{(i,k)}}.
  \label{eq:ProjectorQuadrature}
\end{equation}
Here, the weights $w_k$ and the complex energy shifts $z_k$ of each quadrature node are determined automatically by the integration scheme.
The MPS $\ket{\Psi^{(i,k)}}$ associated with a given node $k$ and guess state $i$ is obtained by solving the linear system:
$(z_k - \mathcal{H}) \ket{\Psi^{(i,k)}} = \ket{\Psi^{(i)}_{\text{guess}}}$.
After every iteration, the Hamiltonian $\mathcal{H}$ is diagonalized within the space spanned by the projected states, yielding approximate eigenpairs of the energy interval $I_E$.
The resulting eigenfunctions serve as updated guesses for the next iteration, and this procedure is repeated until convergence is achieved.
To perform a DMRG[FEAST] calculation, the user must specify a target energy interval by setting $E_{\text{min}}$ and $ E_{\text{max}}$ and the number of $M$ initial guess states, which should be larger than or equal to the number of eigenstates contained in $I_E$.

\paragraph{Vibrational Excited States of the Formic Acid Dimer}

Supramolecular complexes are a challenging class of molecular systems for vibrational calculations.
Due to the weak nature of intermolecular forces, these systems are usually characterized by several highly anharmonic low-frequency modes, which exhibit large amplitude motion.
As a result of these effects, the harmonic approximation commonly used in standard vibrational calculations breaks down. 
Vibrational DMRG (vDMRG) can account for anharmonicities, and 
to demonstrate the QCMaquis vDMRG algorithm, we calculated the ground and several low-lying excited states of the formic acid dimer (FAD), which can be considered a prototypical example for molecular recognition and supramolecular complexes.
For these vDMRG calculations, the analytic PES developed by Qu and Bowman~\cite{qu2016fad_pes} was exploited.
First, the FAD minimum molecular structure was calculated.
The PES minimum is characterized by a planar structure that belongs to the $\text{C}_{2\text{h}}$ point group.
Next, the normal modes and corresponding frequencies were calculated (see supporting information for the computed harmonic frequencies).

In order to perform the vDMRG calculations, the PES around the minimum structure was transformed into a form recognized by QCMaquis using two distinct strategies.
The first strategy consists of the use of $\mathcal{H}_{\text{Watson}}$ from Eq.~(\ref{eq:watson}), for which the Taylor expansion of the PES is needed.
This expansion was constructed by calculating second-, third-, and fourth-order PES derivative tensors around the minimum structure.
The second strategy relies on the $n$-mode Hamiltonian from Eq.~(\ref{eq:hamilton_nmode_first_quant}).
For this strategy, the PES was truncated at the 2-mode coupling terms around the minimum, where each $n$-mode term was represented on an equidistant grid of 31 points in the range between the two 7th harmonic inversion points.
The $n$-mode potentials were then used for the VSCF calculation, in the Fourier discrete variable representation (DVR) basis, performed using the Colibri software.
For the eight lowest-frequency normal modes, the lowest $N_{i}=6$ VSCF modals were used to construct the second quantized form of $\mathcal{H}_\text{nmode}$ from Eq.~(\ref{eq:hamilton_nmode}), while $N_{i}=2$ VSCF modals were used for the other normal modes. This Hamiltonian was then used for the vDMRG calculations.
The low-lying excited vibrational states were computed using both the DMRG[ORTHO] and DMRG[FEAST] algorithms using the single-site variant, with the maximal bond dimension set to $50$.
Examples of the QCMaquis input files for both cases are provided in the supporting information.
The optimized MPS representations of each vibrational state were used to extract the most significant configurations using the SRCAS protocol~\cite{boguslawski2011construction}.
The configuration with the largest weight defines the character of each excited state.
The results are summarized in Table~\ref{tab:fad_dmrg}.
\begin{table}[htb!]
    \begin{tabular}{@{}ccccc@{}}\toprule
            & \multicolumn{3}{c}{Watson} &  \\\hline
         No. & State & $E(\text{ORTHO})$ & $E(\text{FEAST})$ & Harmonic\\\hline
         GS & $0.95\text{GS}_1-0.07\nu_3\nu_5\nu_{22}+0.06\nu_3^2\nu_{21}$ & 15378 & 15380 & 15582\\\hline
         1 & $0.89\nu_1+0.14\nu_3+0.10\nu_1^3$ & 51 & 51 & 71 \\
         2 & $0.81\nu_1^2+0.21\nu_1\nu_3-0.12\text{GS}$ & 104 & 113 & 142 \\
         3 & $0.89\nu_3-0.21\nu_1-0.09\nu_3^2\nu_5\nu_{22}$ & 139 & 139  & 171\\
         4 & $0.94\nu_2-0.07\nu_2\nu_3\nu_5\nu_{22}+0.06\nu_1^2\nu_2$ & 159 & 143 & 167 \\
         5 & $0.77\nu_1^3-0.24\nu_1+0.19\nu_1^2\nu_3$ & 175 & 179 & 213\\
         6 & $0.63\nu_4+0.56\nu_1\nu_3-0.24\nu_1^2$ & 189 & 189 & \multirow{2}{*}{209}\\
         7 & $0.68\nu_4-0.56\nu_1\nu_3+0.09\nu_4^2$ & 199 & 192 \\
         8 & $0.75\nu_5+0.19\nu_1^2\nu_5+0.17\nu_1\nu_3\nu_5$ & 204 & 206 & 254\\
         9 & $0.81\nu_1\nu_2+0.24\nu_5+0.10\nu_1^3\nu_2$ & 208 & 226 & 238 \\
         10 & $0.68\nu_1\nu_4-0.28\nu_1\nu_5-0.22\nu_1^2\nu_3$ & 234 & 236 & 280 \\ \midrule
         & \multicolumn{3}{c}{$n$-mode}  \\\hline
         No. & State & $E(\text{ORTHO})$ & $E(\text{FEAST})$ & Harmonic\\\hline
         GS & $0.99\text{GS}_1-0.07{\nu'}_1{\nu'}_3-0.05{\nu'}_1^2$ & 15456 & 15456 & 15582 \\\hline
         1 & $0.95{\nu'}_1-0.16{\nu'}_3-0.15{\nu'}_1^3$ & 87 & 87 & 71 \\
         2 & $0.86{\nu'}_2-0.11{\nu'}_1{\nu'}_3-0.06{\nu'}_1{\nu'}_2{\nu'}_3$ & 167 & 167 & 167 \\
         3 & $0.77{\nu'}_1^2+0.21{\nu'}_1{\nu'}_3-0.17{\nu'}_1^4$ & 172 & 172 & 142 \\  
         4 & $0.98{\nu'}_4-0.07{\nu'}_1{\nu'}_3{\nu'}_5-0.07{\nu'}_6^2$ & 200 & 200 & 209 \\
         5 & $0.94{\nu'}_3+0.20{\nu'}_1-0.13{\nu'}_1{\nu'}_3^2$ & 224 & 224 & 171 \\
         6 & $0.86{\nu'}_1{\nu'}_2-0.32{\nu'}_1^3-0.15{\nu'}_1^3{\nu'}_2$ & 253 & 253 & 238 \\
         7 & $0.74{\nu'}_1^3+0.38{\nu'}_1{\nu'}_2-0.27{\nu'}_1^2{\nu'}_3$ & 264 & 266 & 213 \\
         8 & $0.96{\nu'}_6-0.17{\nu'}_4{\nu'}_6-0.08{\nu'}_2{\nu'}_6$ & 272 & 272 & 276\\
         9 & $0.93{\nu'}_1{\nu'}_4+0.16{\nu'}_1^3-0.15{\nu'}_3{\nu'}_4$ & 286 & 286 & 280 \\
         10 & $0.98{\nu'}_5+0.07{\nu'}_5^3+0.05{\nu'}_5^2{\nu'}_{11}$ & 291 & 290  & 254\\
         11 & $0.80{\nu'}_1{\nu'}_3-0.32{\nu'}_3^2-0.31{\nu'}_1^2$ & 313 & 304 & 242\\ \bottomrule
    \end{tabular}
    \caption{
        vDMRG energies of low-lying vibrational states of the formic acid dimer calculated using analytic PES from Ref.~\citenum{qu2016fad_pes}.
        $\nu_i^j$ and ${\nu'}_i^j$ denote the $j-$th excited modal of the $i$-th mode for the Watson and $n-$mode Hamiltonian, respectively.
        The ground state energy is given with respect to the PES minimum structure, while the energies of excited states are given relative to the ground state.
        All energies are reported in $\text{cm}^{-1}$.
        In addition to the energies, the three largest CI coefficients, obtained through the SRCAS protocol~\cite{boguslawski2011construction} and the DMRG[ORTHO] MPS, are reported for each vibrational state.
        Each harmonic energy corresponds to the term with the largest CI coefficient. 
    }
    \label{tab:fad_dmrg}
\end{table}
        
For the Watson model, vDMRG reproduces the zero-point energy (ZPE) in excellent agreement with the diffusion Monte Carlo (DMC) value ($15337 \pm 7 \ \text{cm}^{-1}$) reported by Qu and Bowman~\cite{qu2016fad_pes}.
It should be noted that perfect agreement is not expected, as the DMC calculation was performed on the original PES, while the Watson model used approximates the PES by neglecting higher than fourth-order terms in Taylor expansion. For the $n$-mode model, on the other hand, the discrepancy in the ZPE is somewhat larger when compared to the DMC. Both models, however, yield a significant improvement over the harmonic ZPE ($15582 \ \text{cm}^{-1}$). It should be pointed out that both the Watson and the $n$-mode model use approximate potential energy surfaces, obtained by Taylor and $n$-mode expansions respectively, which differ. In fact, the discrepancy of the ZPE in the $n$-mode approach can be traced back to the neglect of the $3$- and $4$-mode terms in the PES representation, which are partially included in the Taylor expansion employed in the Watson model. This is confirmed by running the vDMRG calculation with the Watson model in which $3$- and $4$-mode terms in the third and fourth-order derivative tensors are set to zero (see Supporting information), in which case the ZPE obtained ($15496 \ \text{cm}^{-1}$) closely matches the one from the $n$-mode model.

When the excitation energies obtained with the Watson and the $n$-mode models are compared, the Watson-model vDMRG energies are systematically shifted to lower values, albeit with different magnitudes when compared to the harmonic ones for the first 10 excited states calculated. On the other hand, such a trend is absent in the $n$-mode model, which can again be traced back to the inclusion of the $3$- and $4$-mode coupling terms in the expression of $\mathcal{H}_\text{Watson}$. This fact indicates that the excitation values obtained with the Watson model are expected to be more accurate in the case of the formic acid dimer than the ones obtained with the $n$-mode approach for the employed PES approximations.
Additionally, for the $\nu_4 / {\nu'}_4$ excited states, for which the experimental excitation energy is available ($194 \ \text{cm}^{-1}$), both models yield an improvement over the vibrational configuration interaction ($208 \ \text{cm}^{-1}$)~\cite{qu2016fad_pes}.
However, the assignment of this state in the Watson model is ambiguous, due to the comparable weights of the $\nu_4$ configuration in the $6$-th and $7$-th excited states. As a consequence, they share the same harmonic excitation energy in Table~\ref{tab:fad_dmrg}. When the weights of the most significant configurations of each state are compared between the Watson and the $n$-mode model, it can be noted that their values are significantly larger in the $n$-mode case. This is directly related to the fact that the $n$-mode model uses optimized VSCF modals as the basis set functions, which provide a superior representation of the vibrational wave function when compared to the harmonic oscillator functions. These weights can be used to assign the character of each excited state and provide some insight into the correlation between different modes and different basis functions. Further insight into correlations can be gained from the entanglement diagrams of each state, an example of which is provided in section~\ref{sec:autocas}.

\section{Time-Dependent DMRG}\label{sec:time-dependent}
\subsection{Real-Time Propagation for Quantum Dynamics}
Quantum dynamics calculations are available in QCMaquis via the tangent-space formulation of the time-dependent DMRG (TD-DMRG) algorithm~\cite{Holtz2012Mar,lubich14_timeintegrationtt,haegeman16_mpo-tddmrg}.
In this formulation, the equation for propagating the time-dependent MPS is obtained from the Dirac--Frenkel variational principle
\begin{equation}
  \bra{\delta \Psi} \mathcal{H} - i\partial_t \ket{\Psi} = 0,
  \label{eq:DiracFrenkel}
\end{equation}
where $\ket{\delta \Psi}$ denotes an infinitesimal variation of the wavefunction $\ket{\Psi}$ within the manifold of MPSs of bond dimension $D$.
The resulting equations of motion obtained are of the form
\begin{align}
  \nonumber
  i\partial_t \ket{\Psi_{\text{MPS}}} &= \mathcal{P}_{\Psi_\text{MPS}} \hat{H} \ket{\Psi_{\text{MPS}}} \\
  \ket{\Psi_{\text{MPS}}(t+\Delta t)} &= \text{e}^{-i\Delta t \mathcal{P}_{\Psi_\text{MPS}} \hat{H}} \ket{\Psi_{\text{MPS}}(t)},
  \label{eq:TDDMRG_EOM}
\end{align}
where $\mathcal{P}_{\Psi_{\text{MPS}}}$ denotes the projector onto the tangent space of this manifold with
respect to the reference wavefunction $\ket{\Psi_{\text{MPS}}}$.
Analytical expressions for $\mathcal{P}_{\Psi_{\text{MPS}}}$ can be derived and are found in Refs.~\citenum{lubich14_timeintegrationtt,Holtz2012Apr} and \citenum{haegeman16_mpo-tddmrg}.
Due to the invariance of the MPS from Eq.~(\ref{eq:mps}) with respect to the gauge transformations $\tilde{\mathbf{M}}^{\sigma_k} = \mathbf{G}_{k-1}^{-1} \mathbf{M}^{\sigma_k} \mathbf{G}_k$, the wavefunction $\ket{\Psi_{\text{MPS}}}$ can be transformed to the so-called canonically normalized form with respect to site $k$
\begin{align}
  \nonumber
  \ket{\Psi_{\text{MPS}}} &= \sum_{\boldsymbol{\sigma}} \sum_{\boldsymbol{\alpha}} A_{1\alpha_1}^{\sigma_1}\dots A_{\alpha_{k-2}\alpha_{k-1}}^{\sigma_{k-1}} M_{\alpha_{k-1}\alpha_{k}}^{\sigma_{k}} B_{\alpha_{k}\alpha_{k+1}}^{\sigma_{k+1}} \dots B_{\alpha_{L-1}\alpha_{1}}^{\sigma_{L}} \ket{\sigma_1\dots\sigma_L} \\
  &= \sum_{\sigma_k} \sum_{\alpha_{k-1},\alpha_{k}}  M_{\alpha_{k-1}\alpha_{k}}^{\sigma_{k}} \ket{a^{(l)}_{k-1}\sigma_k a^{(r)}_{k}},
  \label{eq:MPS_k_canonical}
\end{align}
where tensors $\mathbf{A}^{\sigma_j}$ are left-normalized $\sum_{\sigma_j,\alpha_{j-1}} A^{\sigma_j}_{\alpha_{j-1}\alpha_{j}} A^{\sigma_j}_{\alpha_{j-1}\alpha_{j}'} = \delta_{\alpha_{j}\alpha_{j}'}$, while tensors $\mathbf{B}^{\sigma_j}$ are right-normalized $\sum_{\sigma_j,\alpha_{j}} B^{\sigma_j}_{\alpha_{j-1}\alpha_{j}} B^{\sigma_j}_{\alpha_{j-1}'\alpha_{j}} = \delta_{\alpha_{j-1}\alpha_{j-1}'}$.
Using this normalization for the MPS, the projection operator can be expressed as
\begin{align}
  \nonumber
  \mathcal{P}_{\Psi_{\text{MPS}}} &= \sum_{k=1}^L \ket{a^{(l)}_{k-1}\sigma_k a^{(r)}_{k}}\bra{a^{(l)}_{k-1}\sigma_k a^{(r)}_{k}} - \sum_{k=1}^{L-1} \ket{a^{(l)}_{k} a^{(r)}_{k}}\bra{a^{(l)}_{k} a^{(r)}_{k}} \\
  &= \sum_{k=1}^L \mathcal{P}_k^{(1)} - \sum_{k=1}^{L-1} \mathcal{P}_k^{(2)}.
  \label{eq:TDDMRG_projection_operator}
\end{align}
Inserting Eq.~(\ref{eq:TDDMRG_projection_operator}) into (\ref{eq:TDDMRG_EOM}) and approximating the exponential operator using the first order Lie--Trotter splitting, the time evolved wavefunction is given by
\begin{equation}
  \ket{\Psi_{\text{MPS}}(t+\Delta t)} = \text{e}^{-i\Delta t \mathcal{P}^{(1)}_L \hat{H}} \text{e}^{i\Delta t \mathcal{P}^{(2)}_{L-1} \hat{H}}\text{e}^{-i\Delta t \mathcal{P}^{(1)}_{L-1} \hat{H}} \dots \text{e}^{i\Delta t \mathcal{P}^{(2)}_1\hat{H}}\text{e}^{-i\Delta t \mathcal{P}^{(1)}_1 \hat{H}} \ket{\Psi_{\text{MPS}}(t)}.
  \label{eq:TDMPS}
\end{equation}
This expression lends itself to an implementation in a sweep-based DMRG algorithm.
At every site, two local propagation steps are performed: a forward propagation under the action of $\mathcal{P}^{(1)}_k \mathcal{H}$, and a backpropagation step under the action of $\mathcal{P}^{(2)}_k \mathcal{H}$.
Both forward and backward steps are performed with the Lanczos algorithm~\cite{park1986Lanczos_time_evolution,hochbruck1997Lanczos_time_evolution}.
After the local propagation step is completed, a singular value decomposition is applied to the local tensor, and the bond dimension is truncated, as in the standard DMRG approach.

This TD-DMRG algorithm contains several sources of error.
The first source corresponds to the bond dimension truncation, which effectively results in the time evolution under a modified Hamiltonian, as can be seen from Eq.~(\ref{eq:TDDMRG_EOM}).
This error can be reduced by dynamically increasing bond dimension based on the singular values obtained from the truncation step~\cite{baiardi19_td-dmrg}.
The second source of error can be traced to the use of the Lie--Trotter splitting.
In a first-order Trotterization scheme is used, a single sweep across the lattice propagates the MPS to $\ket{\Psi_{\text{MPS}}(t+\Delta t)}$ with an error of the order $\mathcal{O}(\Delta t^2)$.
The error due to Trotterization step can be reduced by adopting a second-order scheme, which sweeps across the entire lattice and back to propagate the MPS to $\ket{\Psi_{\text{MPS}}(t+\Delta t)}$ with the error $\mathcal{O}(\Delta t^3)$.
An additional source of error pertains to the Lanczos algorithm.
In practice, this error can be made arbitrarily small by further enlarging the Krylov space and is usually negligible in comparison to the Trotterization error.

\paragraph{Excitation Transfer in the Benzoic Acid Dimer}
Studying the dynamics of photoinduced processes in molecular dimers and polymers is crucial to understanding the fundamental mechanisms underlying energy and charge transfer in large supramolecular systems~\cite{xie2020perovskite, kim2020donor}.
An important class of such systems is hydrogen-bonded supramolecular complexes, the most prominent example being the nucleobase pairs in the DNA and RNA molecules, whose photodynamics has a significant impact on a plethora of biochemical processes~\cite{mokkath2022dna_exciton_tddft,toppari2013dna_plasmonic,vaya2012dna_excitation_transfer}.
As a prototypical example of such systems, we chose the benzoic acid dimer to study the excitation transfer process in this work.
For this system, the excitation into the $S^{\text{ad}}_2$ excited state can be used to induce the energy transfer process~\cite{ottiger2012excitonic,kalkman2008BAD_splitting_exp}.

\begin{figure}[htb]
    \centering
    \includegraphics[width=\linewidth]{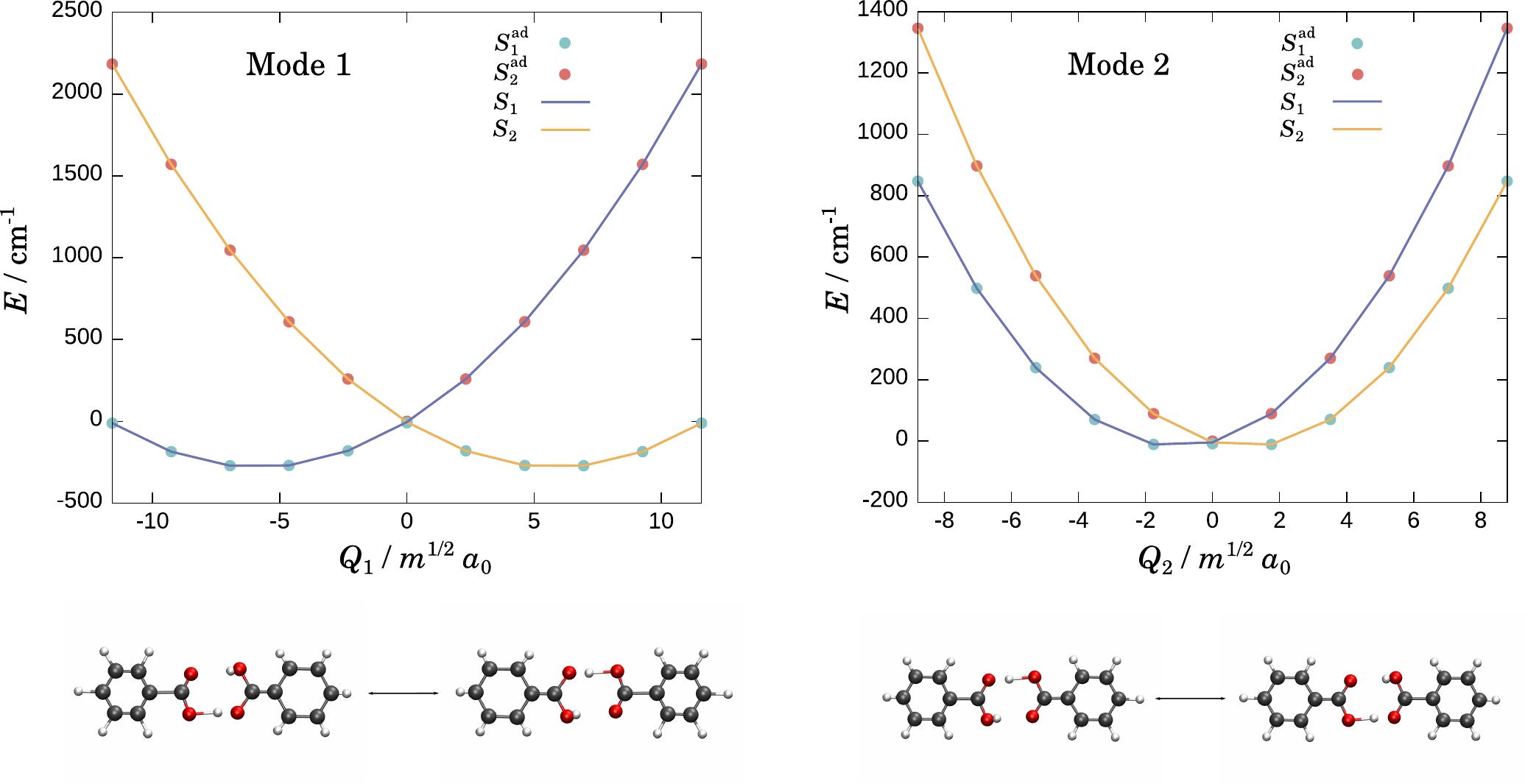}
    \caption{
    PESs of the first two electronic excited states in the benzoic acid dimer along the two normal modes that significantly mediate the energy transfer.
    The scatter and line plots depict the adiabatic and diabatic PESs, respectively.
    Deformations corresponding to the motion along these modes are indicated by the structures at the bottom.
    }
    \label{fig:benzoicacidPES}
\end{figure}

A model of the excitation transfer between the two monomers comprising the benzoic acid dimer can be constructed using only the first two singlet excited states, which, thus, define the vibronic Hamiltonian from Eq.~(\ref{eq:VibronicHamiltonian}).
This process was studied using TD-DMRG to elucidate the population dynamics and determine the vibronic absorption spectrum.
The vibronic Hamiltonian was obtained by Taylor expanding the adiabatic PESs of the first two electronic excited singlet states around the point of minimum energy of the electronic ground state PES pertaining to the $C_{2h}$ point group.
These two adiabatic $S^\text{ad}_1$ and $S^\text{ad}_2$ states correspond to the $A_g$ and $A_u$ irreducible representation (irrep) and arise from a $\pi \to \pi^*$ delocalized excitation on both benzene rings.
At the electronic ground-state equilibrium molecular structure, the $S^\text{ad}_1$ excited state features two imaginary frequencies associated with vibrational modes belonging to the $B_u$ irrep. These modes correspond to antisymmetric ring 'breathing' and ring stretching deformations, as depicted at the bottom of Fig.~\ref{fig:benzoicacidPES}.
Displacements along these modes reduce the system's symmetry from $C_{2h}$ to $C_s$, leading to the localization of the excitation on one of the two benzene rings.
As a result, the $S^\text{ad}_1$ adiabatic state features a double well structure, with each well corresponding to the localization of the excitation on one of the two benzene rings.
The diabatic PESs were constructed following the procedure outlined in Ref.~\cite {raab1999pyrazineVibronicPES}. 
The final model vibronic Hamiltonian employed in this TD-DMRG study includes the 13 vibrational modes that contribute most significantly to the interstate coupling between the $S_1$ and $S_2$ states.

\begin{figure}[htb]
    \centering
    \includegraphics[width=\linewidth]{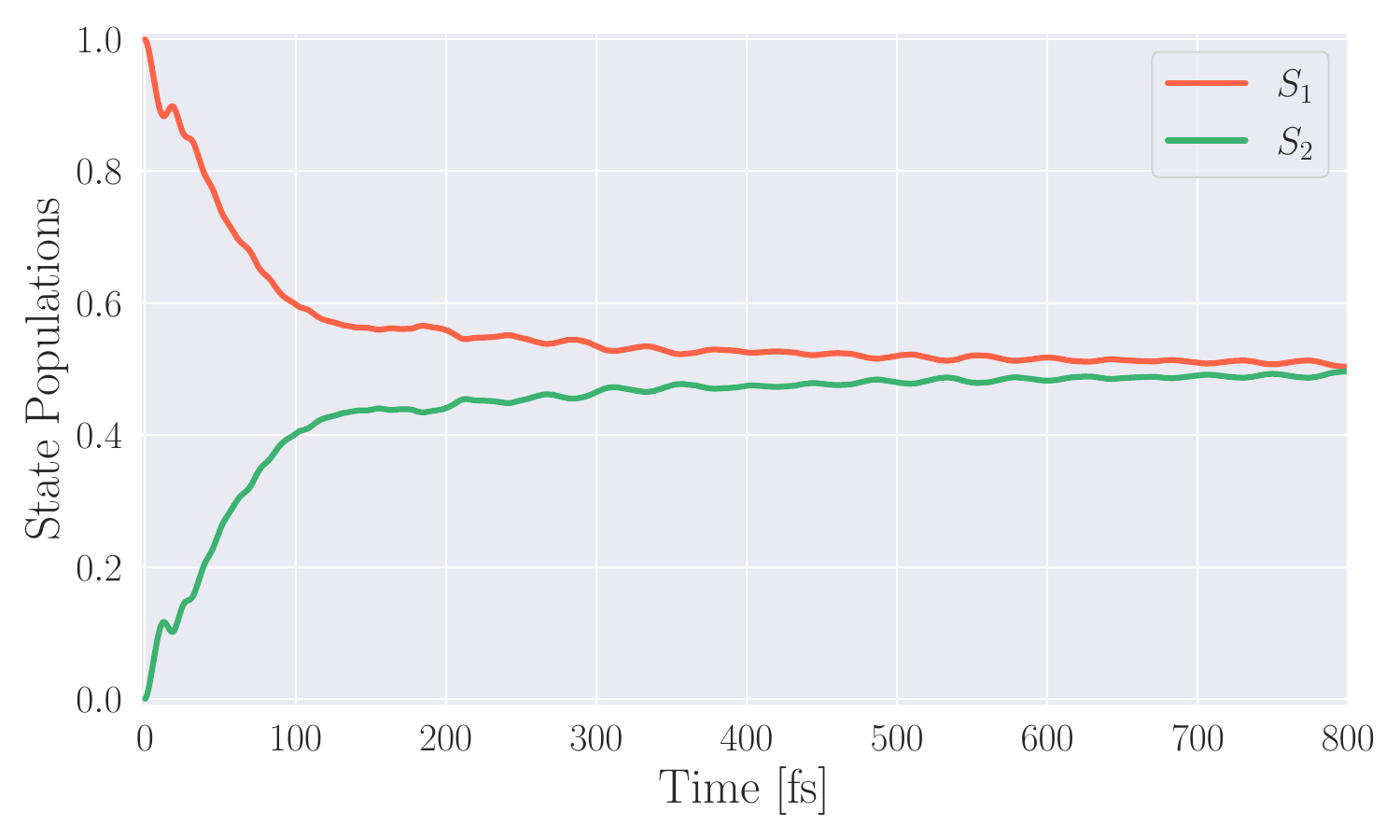}
    \caption{
    Population dynamics for the benzoic acid dimer obtained through TD-DMRG simulations of the vibronic model Hamiltonian.
    The dynamics are initiated from the ground vibrational state of the ground electronic state excited to the $S_1$ electronically excited state.
    }
    \label{fig:populations}
\end{figure}

The study first examined the excitation transfer between the two monomers of the benzoic acid dimer.
The TD-DMRG time propagation was initialized with a state corresponding to the ground vibrational state of the ground electronic state vertically excited to the diabatic $S_1$ electronic state.
The time propagation was executed with a time step $\Delta t = 1\;\text{fs}$ over a total propagation time of 250 fs with a maximum bond dimension of 85.
The population dynamics of the two excited states are depicted in Fig.~\ref{fig:populations}.

During time evolution, the initial excitation confined to the $S_1$ diabatic state migrates to the $S_2$ diabatic state.
After 800 fs, half of the initial excited state population has transferred to the other diabatic states, after which the system approaches equilibrium, with the initial localized excitation being evenly distributed across the two benzene rings of the dimer.
The equilibration timescale obtained from our TD-DMRG calculation is significantly shorter than the exciton transfer rate of 17.7 ps reported in the experimental study of Ref.~\citenum{ottiger2012excitonic}.
This is because we calculated a Franck-Condon excitation directly onto the crossing point of the two diabatic excited state surfaces, where the inter-state coupling is very large.
Ref.~\citenum{ottiger2012excitonic}, on the other hand, reports the exciton transfer rate, defined as the transfer rate of a localized wavepacket in the minimum of one of the diabatic states to the minimum of the other one.

\begin{figure}[htb]
    \centering
    \includegraphics[width=.7\linewidth]{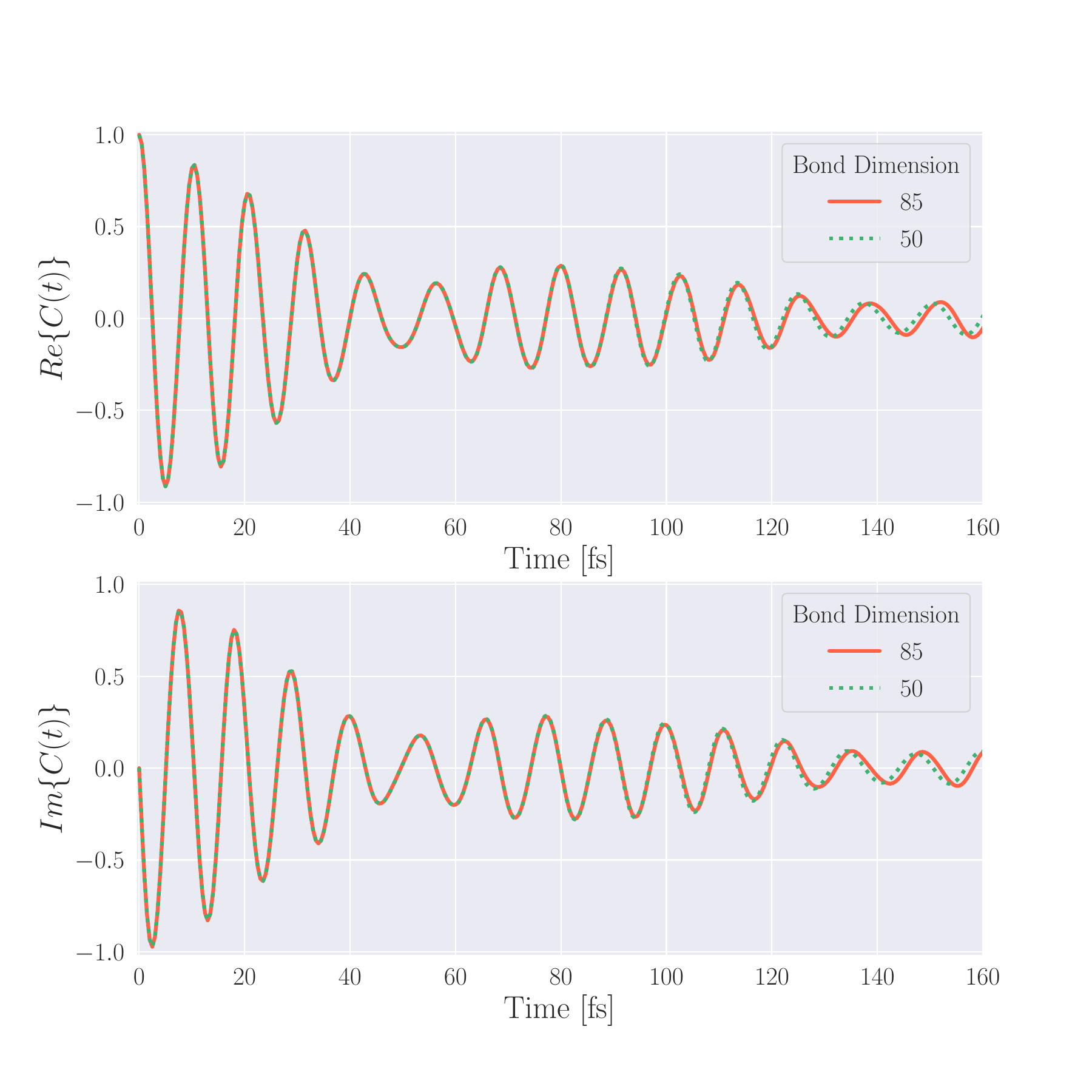}
    \caption{
    Real and imaginary parts of the initial 160 fs of the autocorrelation function for the vibronic model Hamiltonian of the benzoic acid dimer obtained with TD-DMRG.
    The wavepacket is initialized as a coherent superposition of the vibrational ground states of the respective $S_1$ and $S_2$ electronic excited states.
    The simulations were carried out using two different bond dimensions, namely 50 and 85.
    }
    \label{fig:auto}
\end{figure}

Following the investigation of the excitation transfer, the vibronic spectrum of the photoexcitation process was determined.
This spectrum was calculated by performing a Fourier transform of the time-dependent autocorrelation function
\begin{align}\label{eq:autocorr}
C(t) = \braket{\Psi_{\text{MPS}}(0)|\Psi_{\text{MPS}}(t)},
\end{align}
assuming a constant transition dipole moment.
In order to simulate the spectrum, vertical excitation was assumed. For the ground state minimum molecular structure, the $S_2^{\text{ad}}$ is a bright excited state, while the $S_1^{\text{ad}}$ is not. Consequently, to describe the vertical excitation into $S_2^{\text{ad}}$, which is a superposition of $S_1$ and $S_2$ states, the appropriate coherent superposition of the ground state vibrational wavepacket in these two excited electronic states was used as an initial state.
The resulting autocorrelation function is reported in Fig.~\ref{fig:auto}.

\begin{figure}[htb]
    \centering
    \includegraphics[width=\linewidth]{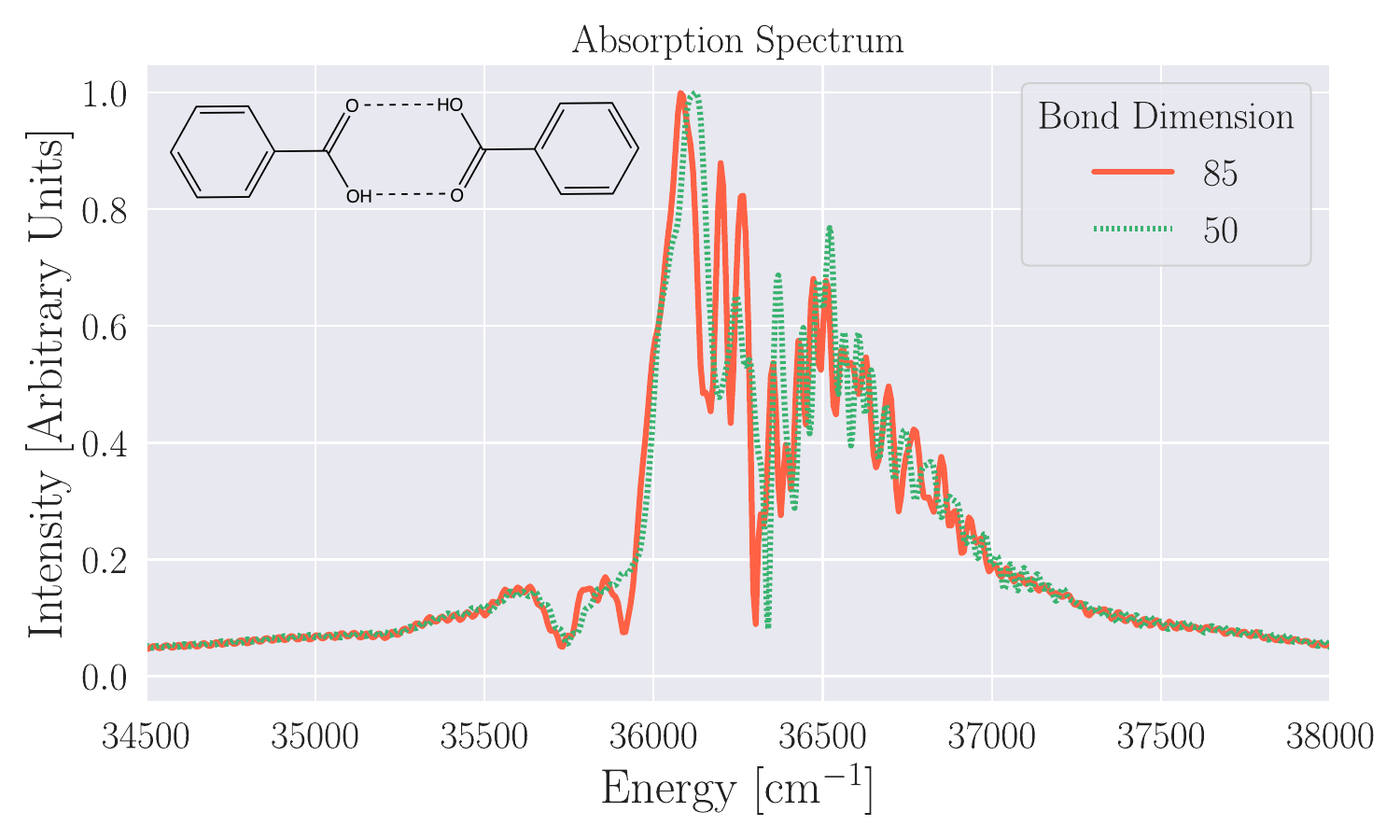}
    \caption{
    Vibronic absorption spectrum of the benzoic acid dimer calculated with TD-DMRG obtained from the autocorrelation function from Fig.~\ref{fig:auto}.
    Intensity is given as the absolute value of the complex spectrum.
    }
    \label{fig:spectrum}
\end{figure}

The autocorrelation function exhibits a slow decay after approximately 80 fs.
When using a lower bond dimension of 50, the initial evolution of the autocorrelation function remains unchanged from the one obtained by the TD-DMRG calculation with a bond dimension of 85. After 120 fs, the data obtained by the two quantum dynamics calculations start to diverge slightly.
The autocorrelation function obtained from these dynamics is reported in Fig.~\ref{fig:spectrum}.

To account for the vertical excitation energy, the spectrum has been shifted by $39752 \; \text{cm}^{-1}$, from which the sum of the zero point energies of all vibrational modes, $6510 \; \text{cm}^{-1}$, was subtracted.
The peaks in the spectrum obtained with a bond dimension of 85 exhibit sharper absorption peaks compared to the one obtained with a bond dimension of 50, illustrating an improvement in accuracy.

\paragraph{Frenkel Exciton Dynamics Across a Rubrene Aggregate}

$\pi$-Conjugated molecular aggregates have garnered significant attention in recent years due to their promising applications in organic electronic devices, such as field-effect transistors and light-emitting diodes~\cite{coropceanu2007charge, smieja2012kinetic}.
A notable example is rubrene single crystals, which demonstrate exceptional hole mobility, positioning them as prime candidates for next-generation field-effect transistors~\cite{takeya2007very}.
Upon photoexcitation, excitons in the rubrene crystal can propagate through electronic coupling, driven by the overlap of the large, delocalized $\pi$-systems of the aromatic monomers.
Given their potential in optoelectronic applications, understanding their interactions with light and the resulting charge transfer dynamics is of particular importance.

\begin{figure}[htb]
    \centering
    \includegraphics[width=0.8\linewidth]{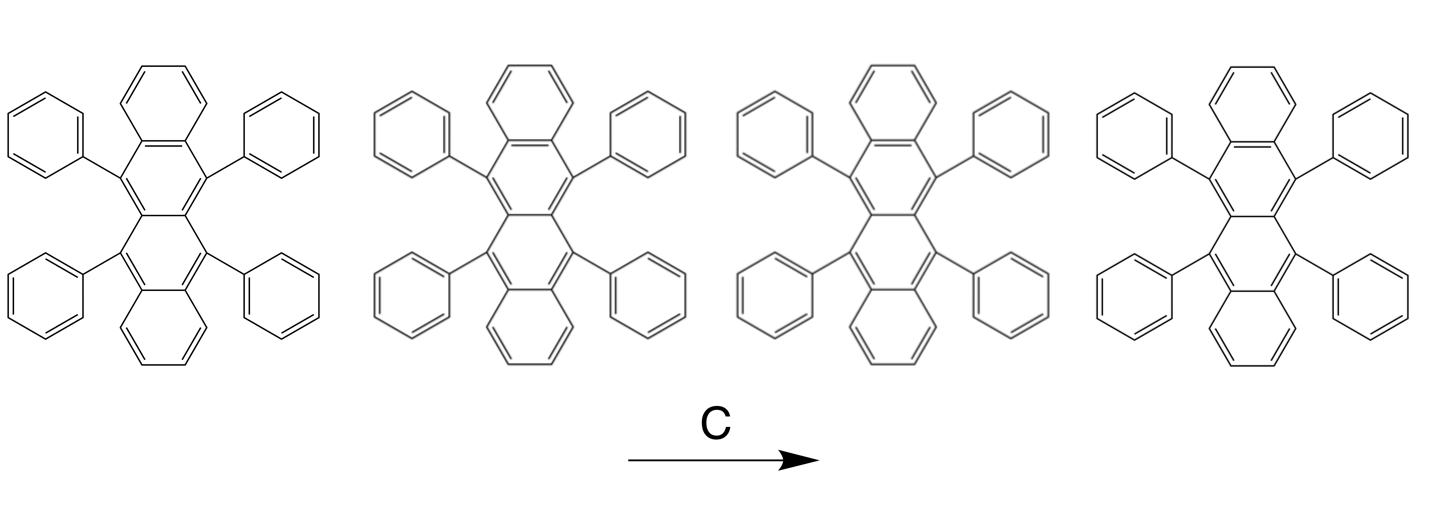}
    \caption{
    A single layer of four rubrene molecules aligned along the c-axis of the rubrene crystal.
    }
    \label{fig:rubrene}
\end{figure}

Here, we investigate the quantum dynamics of an exciton within a single layer of four rubrene molecules aligned along the c-axis of the crystal, which is typically the axis along which charge transfer occurs in these systems.
This tetramer is illustrated in Fig.~\ref{fig:rubrene}.

The Frenkel excitonic Hamiltonian, as described in Section~\ref{sec:vibronic}, was employed to model the behavior of the rubrene crystal upon photoexcitation.
TD-DMRG was used to calculate the absorption spectrum and examine the exciton migration dynamics along the aggregate.

The excitonic Hamiltonian of Eq.~(\ref{eq:hamilton_exciton}) incorporates a constant nearest-neighbor electronic coupling $J_\text{coupl}$, which mediates charge transfer between monomers.
Each monomer is modeled with only two electronic states: the ground state and the first excited state, both represented by harmonic oscillator potentials.
The TD-DMRG calculations were initiated with the first rubrene monomer excited to the vibrational ground state of its first electronic excited state and all other monomers were in their electronic and respective vibrational ground states.
The parameters that enter the excitonic Hamiltonian, consisting of the vibrational frequency, linear shift coefficient $k_i$ from Eq.~(\ref{eq:watson}), and $J_\text{coupl}$-coupling strength, were taken from Ref.~\citenum{gao2009vibrationally}.
The bond dimension was set to 40, and the time step was chosen to be 1 fs with a total propagation time of 1000 fs. 

\begin{figure}[htb]
    \centering
    \includegraphics[width=\linewidth]{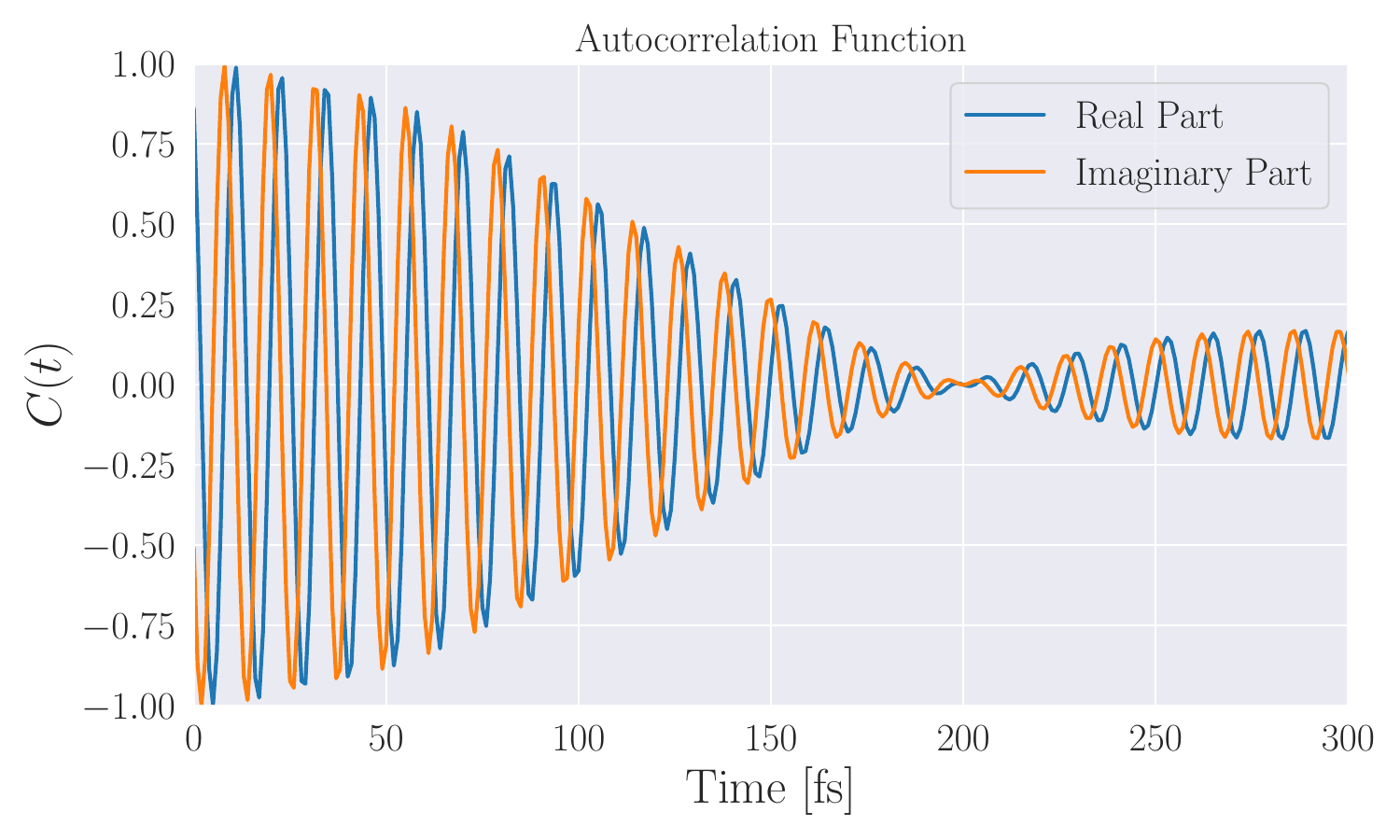}
    \caption{
    Initial 300 fs of the autocorrelation from the TD-DMRG calculation of the excitonic model Hamiltonian for the rubrene tetramer with a bond dimension of 40.
    }
    \label{fig:rubreneAuto}
\end{figure}

\begin{figure}[htb]
    \centering
    \includegraphics[width=\linewidth]{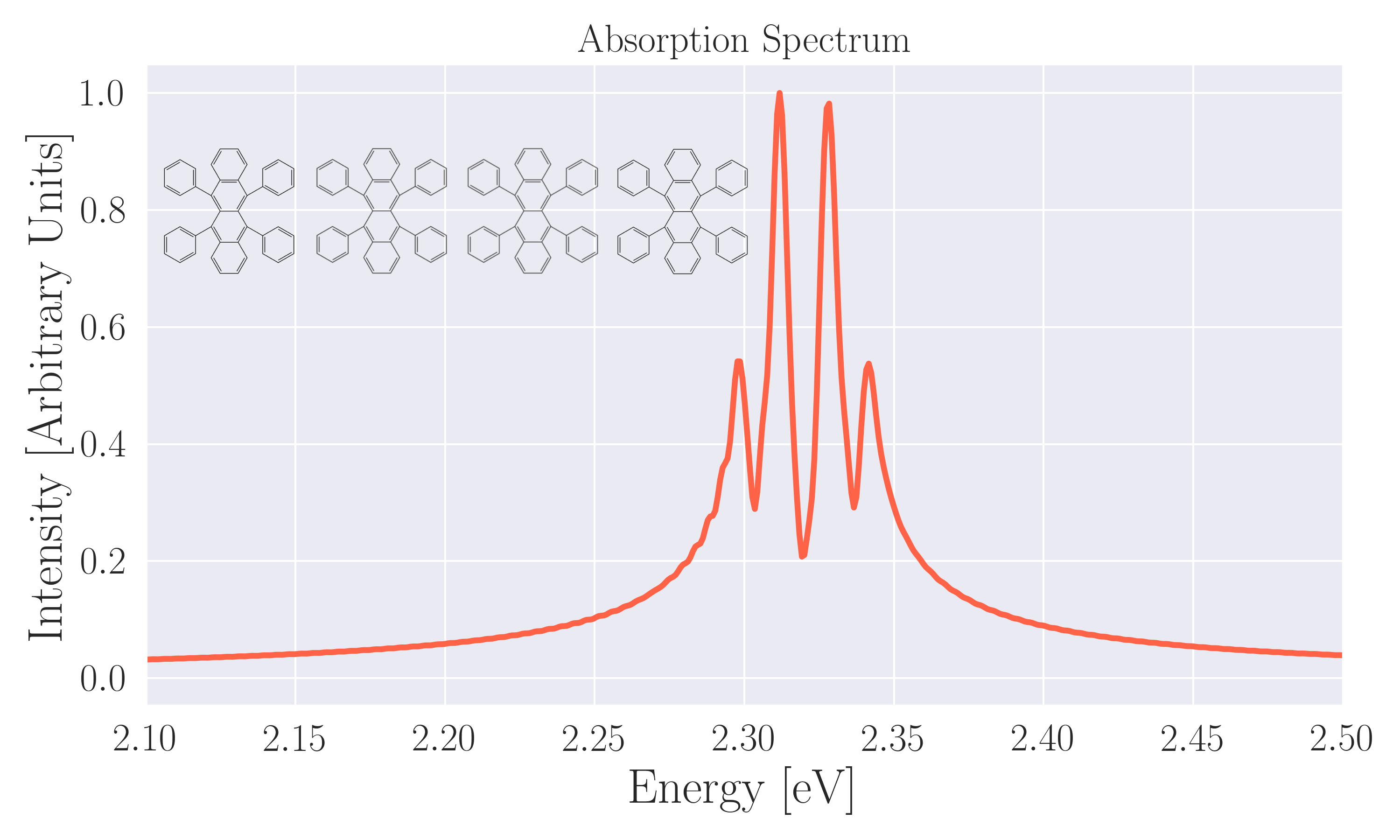}
    \caption{
    Vibronic absorption spectrum of the rubrene tetramer, derived from the autocorrelation function in \ref{fig:rubreneAuto}.
    Intensity is given as the absolute value of the complex-valued spectrum.
    The broad singlet-to-singlet transition appears at approximately $2.32$ eV and exhibits a fine structure arising from the excitonic coupling between the rubrene monomers in the tetramer aggregate.
    }
    \label{fig:rubreneSpec}
\end{figure}

Fig.~\ref{fig:rubreneAuto} shows the autocorrelation function of the first 300 fs of the time evolution.
The autocorrelation function initially decays within the first 200 fs, followed by a revival of its amplitude.
This total autocorrelation function was used to derive the absorption spectrum corresponding to the electronic transition from the ground state to the first excited electronic state.
To reduce spurious oscillatory artifacts in the spectrum caused by finite-time propagation effects, the autocorrelation function is multiplied by an exponential damping factor, $\exp(-t/\tau)$, with $\tau$ = 500 fs, prior to performing the Fourier transform.
The resulting spectrum, shown in Figure~\ref{fig:rubreneSpec}, was shifted to match the experimental 0-0 transition energy of $2.33$ eV.
As expected, the absorption spectrum exhibits vibronic state splitting, which causes a fine structure in the broader peak corresponding to the electronic excitation due to the nearest-neighbor coupling term $J_\text{coupl}$.

\begin{figure}[htb]
    \centering
    \includegraphics[width=\linewidth]{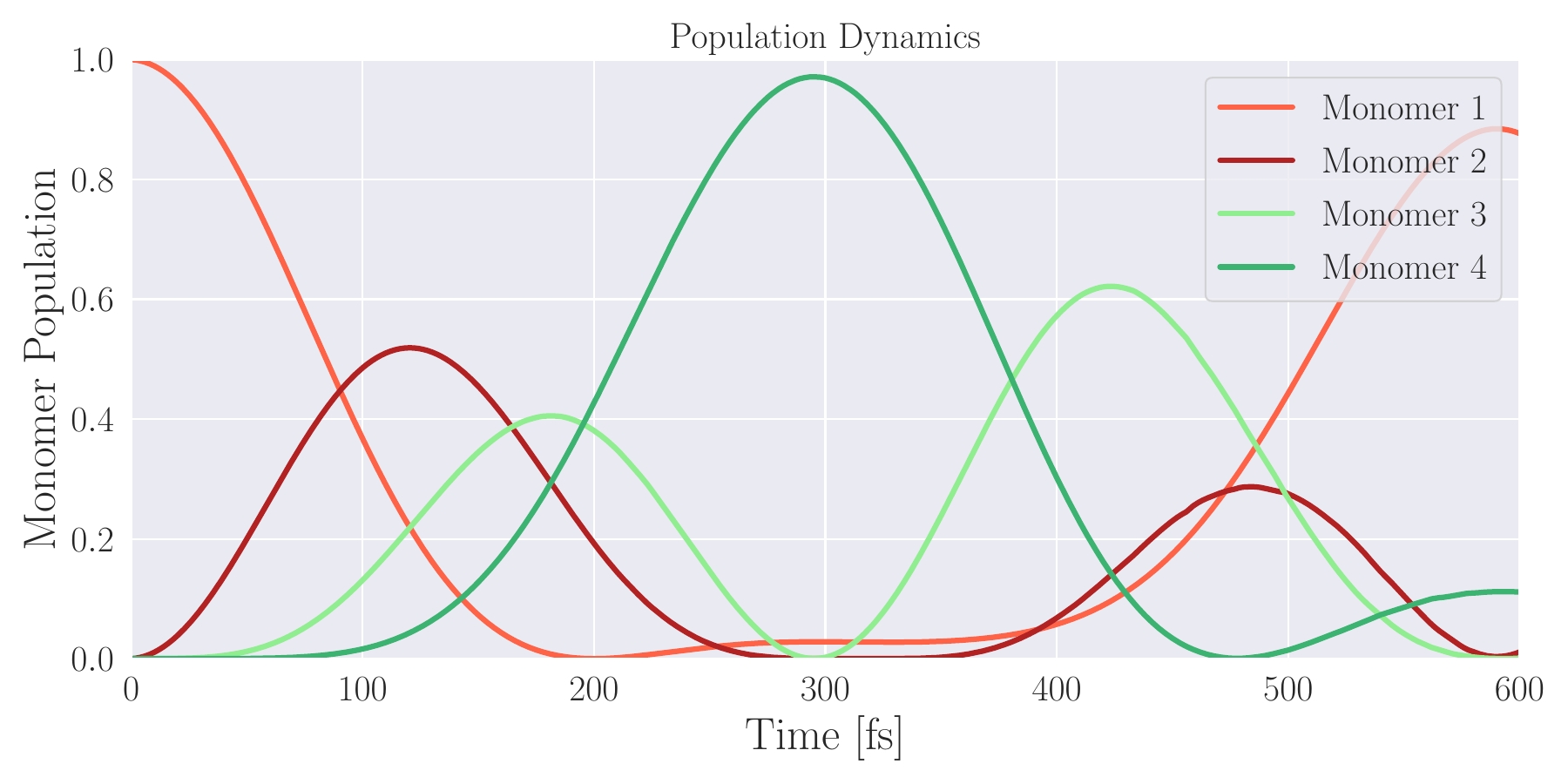}
    \caption{
    Excited-state population dynamics of the excitonic rubrene tetramer.
    This data was obtained through TD-DMRG with a bond dimension of 40.
    }
    \label{fig:rubrenePop}
\end{figure}

The excited-state population dynamics of the rubrene tetramer are presented in Fig.~\ref{fig:rubrenePop}.
Initially localized on the first monomer, the excitation is fully transferred to the other monomers within the first 200 fs of the propagation, coinciding with the vanishing of the autocorrelation function in Fig.~\ref{fig:rubreneAuto}.
As expected, this transfer proceeds sequentially along the chain of monomers, from one neighboring monomer to the next.
A notable accumulation of excited-state population is observed on the fourth monomer after 300 fs.
This behavior is attributed to its position at the terminal end of the tetramer, possessing only a single neighboring monomer for propagating the exciton further.

\subsection{Imaginary-Time Propagation for Ground State Optimization}

The time-dependent formulation of DMRG also provides an alternative method for optimizing the right eigenvectors of the Hamiltonian by performing imaginary-time evolution, expressed as
\begin{align}
  \ket{\Psi} = \lim_{t \rightarrow \infty} e^{-\mathcal{H} t} \ket{\Psi_{\text{initial}}},
\end{align}
starting from an arbitrary initial vector $\ket{\Psi_{\text{intial}}}$.
This evolution converges to the lowest eigenstate that has non-vanishing overlap with the initial state $\ket{\Psi_{\text{initial}}}$.
Provided that the initial wavefunction contains contributions from the ground state, the imaginary-time evolution can effectively be used to find the ground state of an arbitrary Hamiltonian.

\begin{figure}[htb]
    \centering
    \includegraphics[width=0.9\linewidth]{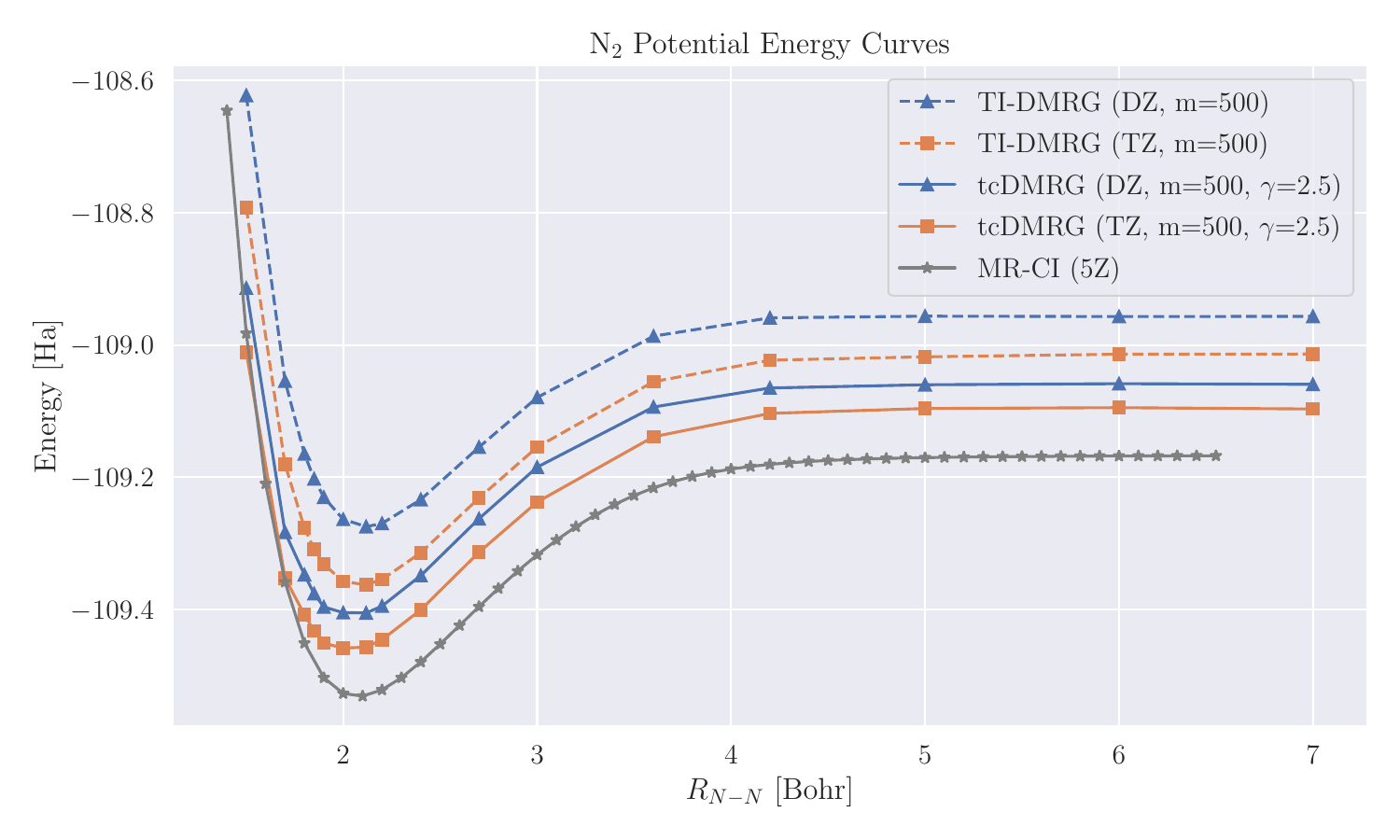}
    \caption{
    N$_2$ potential energy curves for time-independent (TI-) and transcorrelated (tc) DMRG in the cc-pV(DZ) and cc-pV(TZ) basis sets using a bond dimension $m = 500$.
    The tcDMRG calculations were conducted using a correlation factor $\tau = \frac{1}{2}\sum_{i < j}\exp(-\gamma r_{ij})$ with $\gamma=2.5$.
    As a high-accuracy reference curve, the MRCI+Q-F12 in the aug-cc-pV(5Z) basis is taken from the literature~\cite{Gdanitz1998Feb}.
    }
    \label{fig:tcn2-pec}
\end{figure}

As demonstrated in our previous works~\cite{baiardi20_tcdmrg,baiardiExplicitlyCorrelatedElectronic2022}, this optimization method is especially useful for non-Hermitian Hamiltonians, such as the transcorrelated Hamiltonian, where conventional DMRG optimization schemes fail due to the breakdown of the variational principle for non-Hermitian operators.
Fig.~\ref{fig:tcn2-pec} depicts the potential energy curves of the N$_2$ molecule obtained through time-independent and imaginary-time transcorrelated DMRG calculations using the cc-pVDZ and cc-pVTZ bases~\cite{Dunning1989Jan} with a bond dimension of 500.

The correlator used for transcorrelation is given by $F = e^\tau$, where $\tau = \frac{1}{2} \sum_{i < j} \exp(-\gamma r_{ij})$, which exactly enforces the electronic cusp conditions when two electrons coalesce.
The correlation factor of $\gamma$ was set to 2.5 because lower values of $\gamma$ (e.g., $\gamma = 1$) lead to significantly non-variational energies -- see Ref.~\citenum{szenesStrikingRightBalance2024} for further discussion on this issue.

\section{Dynamic Electron Correlation}\label{sec:dynamic-corr}

In practical electronic structure calculations, DMRG is typically employed as an active-space solver, where a subset of the orbitals designated as active form a subspace in which DMRG is executed.
For quantitatively accurate results, the missing dynamic correlation from the neglected orbitals---absent from the active space---can be addressed using post-CAS methods.
Examples of such methods combined with DMRG have been reviewed in Ref.~\citenum{Baiardi2020Jan,Cheng2022Jan}.
Several forms of multi-reference perturbation theory (MRPT) are readily available thanks to our interface to external quantum chemistry programs and are described below.
We note in passing that we have also investigated other approaches within the QCMaquis environment for capturing the dynamical correlation lacking in the DMRG ansatz including range-separated DFT~\cite{Hedegard2015Jun} and multi-reference driven coupled cluster~\cite{Morchen2020Dec,Feldmann2024Oct}.

\subsection{Multi-Reference Perturbation Theory}

QCMaquis provides routines to evaluate up to four-body RDMs derived from an MPS wavefunction.
These RDMs are needed for MRPT methods such as CASPT2~\cite{anderssonSecondorderPerturbationTheory1990,Andersson1992Jan} and N-electron valence state second-order perturbation theory (NEVPT2)~\cite{angeliIntroductionElectronValence2001}.
Our NEVPT2 implementation, which leverages Cholesky-decomposed two-electron repulsion integrals~\cite{freitag17_dmrg-nevpt2}, is available as a separate module in OpenMolcas, while our new CASPT2 implementation is integrated within the conventional OpenMolcas DMRG interface.

In the CASPT2 formalism, computing the correction to the energy involves expressions~\cite{Lindh2020Nov} containing contractions between the RDMs $\Gamma_{pq\ldots}^{st\dots}$ of the CAS wavefunction and the generalized Fock matrix $f_{v,v'}$, such as
\begin{equation}\label{eq:fock-contraction}
    A_{pqr}^{stu} = \sum_{vv'} f_{vv'} \Gamma_{pqrv}^{stuv'} = \sum_{vv'} f_{vv'} \braket{\psi | e_{pqrv}^{stuv'} |\psi},
\end{equation}
with $e_{pq\dots}^{st\ldots}$ corresponding to the $n$-electron spin-summed excitation operators~\cite{Helgaker2000Aug}.
Eq.~(\ref{eq:fock-contraction}) requires computing the 4-RDM, which will become prohibitively memory-intensive for large active spaces where exact diagonalization methods are infeasible and approximate CAS solvers, such as DMRG, become necessary. 
To make our CASPT2 implementation practical, certain simplifications are made to Eq.~(\ref{eq:fock-contraction})~\cite{Kurashige2011Sep}.
First, the molecular orbitals are rotated to pseudo-canonical form, which diagonalizes $f_{v,v'}$, eliminating one of the indices in the sum.
Furthermore, the four-body excitation operator is expressed in terms of lower-body excitation operators as follows:
\begin{equation}
 e_{pqrv}^{stuv} = e_{pqr}^{stu}E_{vv} - e_{pqr}^{stv}\delta_{vu} - e_{pqr}^{svu}\delta_{vt} - e_{pqr}^{vtu}\delta_{vs}.
\end{equation}
Then, the contraction in Eq.~(\ref{eq:fock-contraction}) can be rewritten as:
\begin{align}\label{eq:fock-simplified}\notag
A_{pqr}^{stu} &= \sum_{v}f_{vv}  \Gamma_{pqrv}^{stuv}  \\\notag
&= \sum_{v}f_{vv}\left(\braket{ \psi | e_{pqr}^{stu}E_{vv} | \psi } - \delta_{vu}\Gamma_{pqr}^{stv} - \delta_{vt}\Gamma_{pqr}^{svt} - \delta_{vs}\Gamma _{pqr}^{vtu}\right) \\\notag
&= \braket{ \psi | e_{pqr}^{stu} \sum_{v}f_{vv}E_{vv}|\psi } -f_{uu}\Gamma_{pqr}^{stu} - f_{tt}\Gamma_{pqr}^{stu} - f_{ss}\Gamma_{pqr}^{stu} \\
&= \braket{ \psi | e_{pqr}^{stu} |\psi' } - \Gamma_{pqr}^{stu} (f_{uu} + f_{tt} + f_{ss}).
\end{align}
Here, $\ket{\psi'}$ corresponds to a new wavefunction resulting from the application of the operator $\hat{O} = \sum_v f_{vv} E_{vv}$ on the original wavefunction $\hat{O}\ket{\psi} = \ket{\psi'}$.
In the DMRG formalism, this operation is given by the contraction between an MPO and an MPS.
Although this operation can, in principle, be expensive, in this case, it can be performed efficiently because the operator $\hat{O}$ takes the form of an MPO of bond dimension of only 2 due to its diagonal nature.
The first term in Eq.~(\ref{eq:fock-simplified}), therefore, denotes a transition 3-RDM between the newly computed MPS $\ket{\psi'}$ and the original one $\ket{\psi}$.
With this approach, all quantities in the CASPT2 algorithm requiring the 4-RDM may be derived from the 3-RDM and this transition 3-RDM, eliminating the need to explicitly compute and store the expensive 4-RDM.
To further accelerate calculations, an option is provided in OpenMolcas to compress the MPS to a smaller bond dimension before evaluating the transition 3-RDM, which is the bottleneck of the computation.
While this option has the potential to significantly speed up the RDM evaluation, excessive compression can compromise the accuracy of the results.
Our current CASPT2 implementation is limited to state-specific calculations with a multi-state extension under development.

\begin{figure}[htb]
    \centering
    \includegraphics[width=\linewidth]{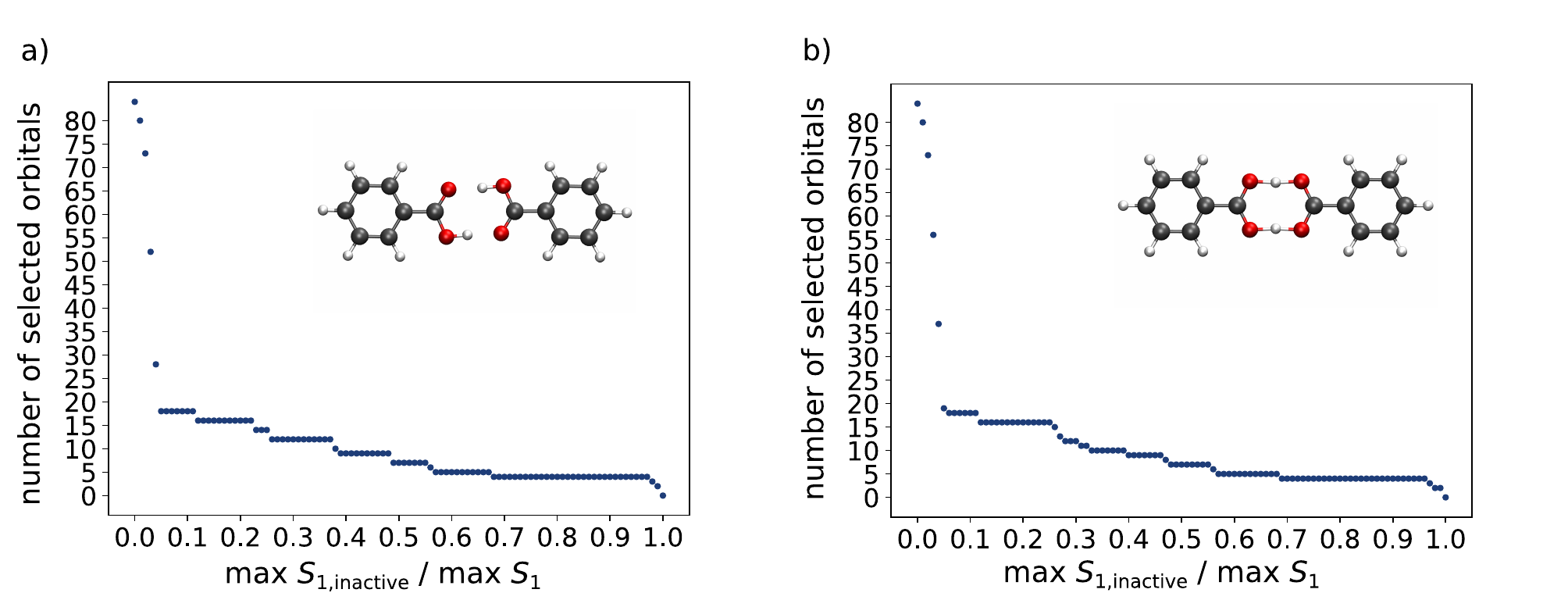}
    \caption{
        Threshold diagrams corresponding to the maximal discarded single-orbital entropies relative to the largest value for different active space sizes, as introduced in Ref.~\citenum{stein16_automatedselection} for a) minimum structure and b) transition state structure of the benzoic acid dimer.
    }
    \label{fig:threshold_bad}
\end{figure}

\paragraph{Minimum and Transition State Energy Barrier of the Benzoic Acid Dimer}
To demonstrate the importance of dynamic correlation effects in this context, we investigate the electronic energy barrier between the minimum and transition state structures of the benzoic acid dimer.
The transition state structure corresponds to the $\text{D}_{2\text{h}}$ point group, with both carboxyl hydrogen atoms equidistant to the corresponding oxygen atoms of both monomers. In the minimum structure, on the other hand, each carboxyl hydrogen is associated with one monomer, which reduces the symmetry of the molecule to $\text{C}_{2\text{h}}$.
For both geometries, the active space of (20e, 18o) was determined by AutoCAS~\cite{stein16_automatedselection} (see also below), which included the entire $\pi$-system of both monomers in the active space.
For the transition state, this corresponds to the 6 energetically lowest orbitals of the $A_g$ and $B_{3u}$ symmetries, and 3 energetically lowest orbitals of the $B_{2u}$ and $B_{1g}$ symmetries.
In the case of the minimum structure, the selected active space corresponds to the 9 energetically lowest orbitals of the $A_g$ and $B_u$ symmetries.
Threshold diagrams of both geometries are given in Fig.~\ref{fig:threshold_bad}.

\begin{figure}[htb]
    \centering
    \includegraphics[width=0.7\linewidth]{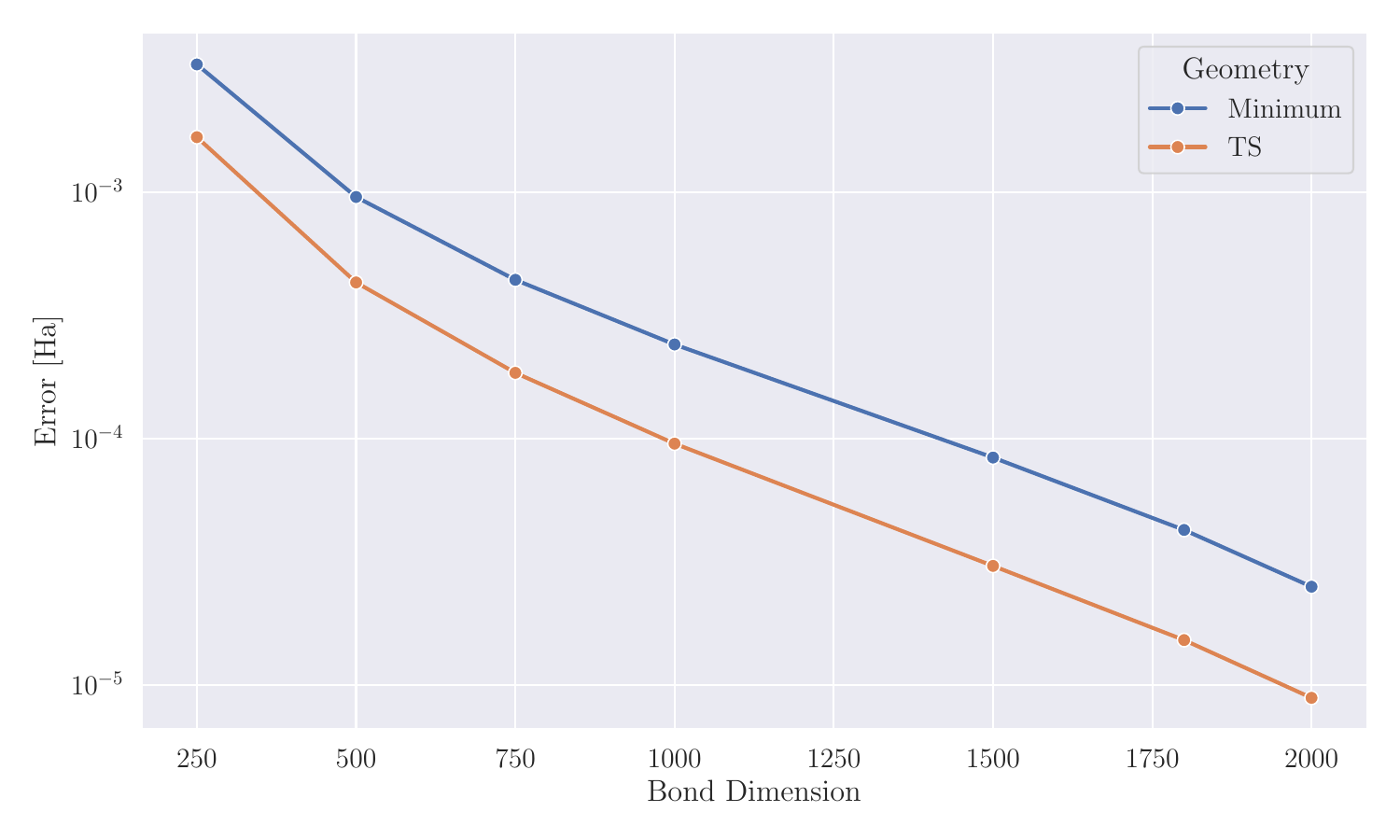}
    \caption{
        Logarithmic plot of the error of the DMRG-SCF electronic energy of the benzoic acid dimer for different bond dimensions measured with respect to the energy obtained with a bond dimension of 2500.
        This error is reported for the minimum and transition state (TS) molecular geometries.
    }
    \label{fig:dmrgscf}
\end{figure}

An initial DMRG-SCF calculation
was performed using the OpenMolcas interface for both the minimum and transition state molecular geometries with the AutoCAS selected (20e, 18o) active space using the def2-SVP basis set~\cite{weigend2005def2-tzvp}.
The orbital mapping was chosen based on the Fiedler vector of the mutual information of the orbitals~\cite{Barcza2011Jan}.
The convergence of the energy with respect to the bond dimension is illustrated in
Fig.~\ref{fig:dmrgscf}.

Fig.~\ref{fig:dmrgscf} depicts the error of the electronic energy obtained for various bond dimensions, measured with respect to the energy with a calculation performed at a bond dimension of 2500.
To reduce the computational cost of evaluating the RDMs for the CASPT2 method, the CASPT2 calculations were performed using the DMRG-SCF wavefunction of bond dimensions 1000.
This choice is deemed justified since both the ground and transition states have an error below $10^{-3}$ [Ha].
The converged MPS was compressed to a bond dimension of 500 for the evaluation of the 3-RDM.
The CASPT2 calculation utilized an empirical ionization-potential–electron-affinity shift~\cite{Ghigo2004Sep} of 0.25 and an imaginary level shift~\cite{Forsberg1997Aug} of 0.1.
The results are summarized in Table~\ref{tab:caspt2}.

\begin{table}[htb]
    \begin{tabular}{@{}ccccc@{}}\toprule
         & DMRG-SCF & DMRG-CASPT2 & DLPNO-CCSD(T) & Experiment \\\midrule
         Minimum & -6236.9 & -8543.6 & -8754.0 & \\
         Transition State & -6204.0 & -8520.1 & -8740.6 &  \\ 
         Barrier & 32.9& 23.5 & 13.4 & 28.4$^a$ \\\bottomrule
         {\footnotesize $^a$ Ref. \citenum{kalkman2008BAD_splitting_exp}} 
    \end{tabular}
    \caption{
        Electronic energies in mHa (absolute energies shifted by $-8.3 \cdot 10^{6}$) of the benzoic acid dimer obtained from a DMRG-SCF and subsequent CASPT2 calculation for the minimum and transition state geometries using an active space of CAS(20e, 18o).
        Both calculations were performed using DMRG with a bond dimension of 1000, and the evaluation of the transition 3-RDM was conducted using an MPS that was compressed to a bond dimension of 500.
    }
    \label{tab:caspt2}
\end{table}

For comparison, DLPNO-CCSD(T) calculations were carried out using the ORCA quantum chemistry package~\cite{Neese2020Jun}.

\section{Interfaces to External Quantum Chemistry Packages}\label{sec:interfaces}

The electronic structure model of QCMaquis is tightly integrated with several quantum chemistry packages, including the newly added support for PySCF.

\subsection{OpenMolcas}

Historically, QCMaquis has served as the default DMRG approximate FCI solver in the OpenMolcas quantum chemistry package, which specializes in methods designed to tackle problems requiring a multi-configurational description of the wavefunction.
As OpenMolcas is written in the Fortran programming language, QCMaquis provides a C interface that exposes routines to its core functionality.
This C interface bridges QCMaquis and OpenMolcas by binding its routines using Fortran functions, allowing direct invocation from within the OpenMolcas source code.

\subsection{Python Bindings and PySCF}

To enhance the standalone usability of the software, Python bindings were developed that expose QCMaquis's functionality.
As Python is increasingly dominant in scientific computing, these bindings facilitate the integration of QCMaquis into custom workflows and improve compatibility with other tools in the Python ecosystem.
The Python bindings are accessed indirectly via concise wrapper functions, which ensure that the internal mechanics of QCMaquis remain abstracted from the user.

Leveraging the Python bindings, QCMaquis provides a newly developed interface to PySCF, enabling its use as an FCI solver.
This allows CASCI and CASSCF calculations to be performed with QCMaquis as the active-space solver by replacing the corresponding \texttt{FCISolver} object.

\begin{figure}[htb]
\begin{lstlisting}[language=Python]
from pyscf import gto, scf, mcscf
from scine_qcmaquis import DMRGSolver

mol = gto.M(atom="N 0 0 0; N 0 0 2.5")
mf = scf.RHF(mol).run() # Hartree-Fock

ncas = 6; nelec = 6
mc = mcscf.CASSCF(mf, ncas, nelec) # or mcscf.CASSCI(mf, ncas, nelec)
mc.fcisolver = DMRGSolver(mol)
e_dmrgci = mc.kernel()[0]
\end{lstlisting}
\caption{
Code snippet of the DMRGCI and DMRGSCF calculation on the stretched N$_2$ molecule with a CAS of (6e, 6o) using the PySCF interface.
}
\end{figure}

\section{Measurements and Properties}\label{sec:measurements}

\subsection{CI coefficients}

In a DMRG calculation, since the wavefunction is represented as an MPS tensor factorization, the CI coefficients are not directly accessible.
However, specific coefficients may be extracted from the wavefunction by evaluating the overlap between the optimized MPS wavefunction $\ket{\Psi_{\text{MPS}}}$ and an MPS corresponding to a single configuration
\begin{equation}\label{eq:ci-coeff}
    c_{p} = \braket{\Psi_{p} | \Psi_{\text{MPS}}},
\end{equation}
where $c_{p}$ corresponds to an arbitrary CI coefficient related to configuration $p$.
In QCMaquis, the individual configuration can be specified using the ONV string of the configuration.

To recover the most significant CI coefficients, our sampling reconstruction of the complete active space (SRCAS) procedure~\cite{boguslawski2011construction} may be employed.
In the SRCAS algorithm, the configurational space is sampled with a Metropolis--Hastings Markov chain.
Several parameters affect the sampling procedure.
All of them are provided with sane default values; however, they can be tuned, if necessary.
These parameters are: the threshold for the completeness of the CI representation $\eta_\text{complete}$, which represents the norm of the reconstructed CI wavefunction and is used to terminate the algorithm (default value is set to $0.99$), and the threshold for storing the sampled ONV $\eta_\text{store}$ (default value is set to $0.001$).
The algorithm consists of the following steps:
\begin{enumerate}
  \item \textbf{Intial Setup:}
  An initial guess ONV is automatically created or can alternatively be provided by the user.
  Its CI coefficient $C_{\text{curr}}$ is computed by evaluating the overlap with the MPS as in Eq.~(\ref{eq:ci-coeff}).
  \item \textbf{Generate New State:}
  From the current reference ONV, a randomly (de)excited state is generated.
  
  \item \textbf{Evaluate New Configuration:} Calculate the CI coefficient $C_{\text{new}}$ of the newly generated ONV, and store the configuration if $\left| C_{\text{new}} \right| > \eta_{\text{store}}$.
  \item \textbf{Update the Reference ONV:} Update the reference ONV with the newly generated one with a probability $P= \min \left[ 1, \frac{\left| C_{\text{new}} \right|^2}{\left| C_{\text{curr}} \right|^2} \right]$.
  \item \textbf{Repeat Until Convergence:} Repeat steps 2 to 5 until the CI expansion is sufficiently reconstructed as measured by $\sum_i \left| C_i \right|^2 > \eta_{\text{complete}}$.
\end{enumerate}

\subsection{Particle Reduced Density Matrices and Transition Particle Densities}

QCMaquis provides routines for computing up to 4-particle RDMs, which are essential for various applications such as orbital rotations (requiring 1- and 2-RDMs) during DMRG-SCF iterations, as well as for certain dynamical correlation methods that depend on additional 3- and 4-RDMs, such as CASPT2.
Users have the flexibility to specify a sub-block for the RDM computation, rather than calculating the entire matrix. The syntax for this option is detailed in the manual.
Additionally, QCMaquis can compute up to 3-particle transition-RDMs between two wavefunctions stored as MPS in the form of checkpoint files.
This functionality is exposed either through the standard input file or the newly developed Python interface.

\begin{figure}[htb]
\begin{lstlisting}[language=Python]
from scine_qcmaquis import QCMaquis

dmrg = QCMaquis()
# ...
dmrg.run(...)
# Returns numpy arrays
rdm1 = dmrg.get_one_rdm();
rdm2 = dmrg.get_two_rdm();
\end{lstlisting}
\caption{
Code snippet for extracting up to 4-body RDMs from QCMaquis.
}
\end{figure}

\subsection{Orbital Entropies and Quantum Entanglement Measures}

QCMaquis provides routines for extracting quantum information metrics, such as the orbital entropies\cite{Legeza2003, Legeza2006} and mutual information\cite{Rissler2006}, from MPS wavefunctions. 
These quantities rely on the evaluation of the orbital RDMs.
Due to the complexity of evaluating certain operator expectation values entering the orbital RDM within a spin-adapted framework, only a subset of them are directly implemented.
Internally, the remaining values are computed by first transforming the MPS to the 2U(1) symmetry.
This transformation converts the spin-adapted MPS to the one with only conservation of $\alpha$ and $\beta$ particles.

These orbital-RDMs can be used to compute the single-orbital entropy for spatial orbital $i$, given by:
\begin{equation}
    s_i(1) = -\sum_{\alpha=1}^4 \omega_{\alpha, i} \ln \omega_{\alpha, i},
\end{equation}
where $\alpha$ runs over the possible occupancies of the orbital and $\omega_{\alpha, i}$ correspond to the eigenvalues of the orbital RDM.
The two-orbital entropies are defined analogously by 
\begin{equation}
    s_{ij}(2) = -\sum_{\alpha=1}^{16} \omega_{\alpha, ij} \ln \omega_{\alpha, ij},
\end{equation}
where $\alpha$ now represents the 16 possible occupancies of two spatial orbitals and $\omega_{\alpha, ij}$ denotes the eigenvalues of the two-orbital RDM for orbitals $i$ and $j$.
These quantities are used to construct the mutual information matrix of the system
\begin{equation}
    I_{ij} = \frac{1}{2} \left(s_1(1) + s_j(1) - s_{ij}(2)\right)(1 - \delta_{ij}).
\end{equation}

\begin{figure}[htb]
    \centering
    \includegraphics[width=0.7\linewidth]{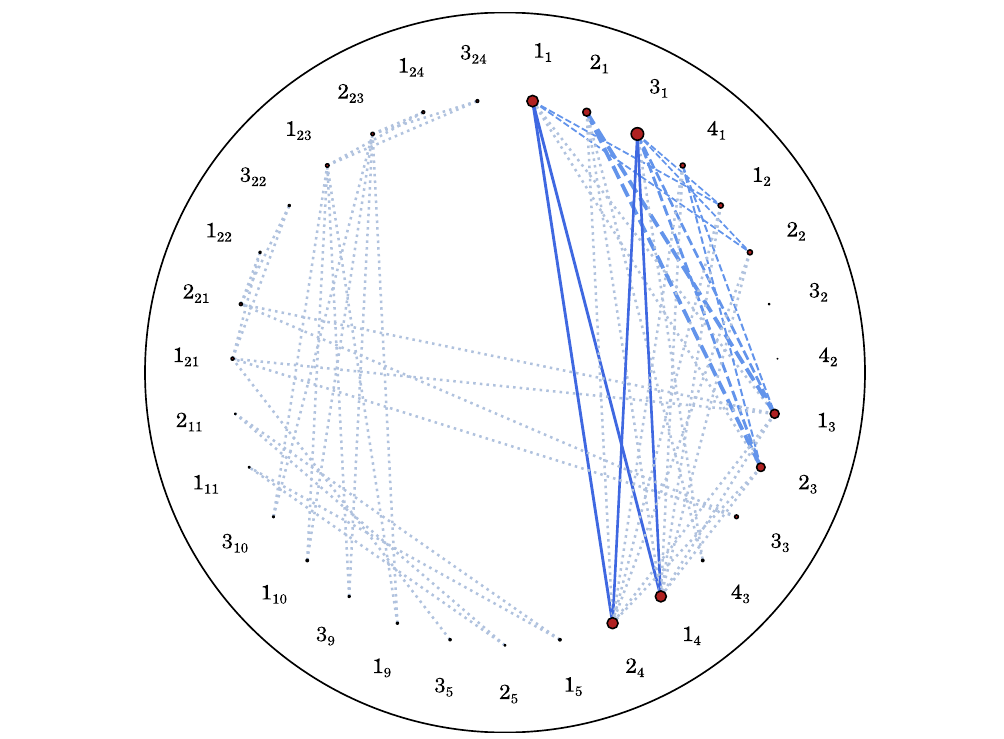}
    \caption{
    Modal entanglement diagram for the vibrational excited state ${\nu'}_1^2$ of the formic acid dimer obtained from the $n$-mode Hamiltonian.
    Only the most entangled modals are presented.
    The symbol $K_i$ denotes $K$-th VSCF modal of the normal mode $i$.
    Circle diameters are proportional to the value of the corresponding modal entropy, while line widths are proportional to the value of the mutual information.
    Further details on the entanglement diagrams can be found in Ref.~\citenum{glaser24_vib-ent}.
    }
    \label{fig:entanglement_fad}
\end{figure}

The entropies and mutual information are also available for vibrational Hamiltonians expressed in the $n$-mode second quantization framework, as introduced in Eq.~(\ref{eq:hamilton_nmode}).
For the corresponding expressions of the single-modal and two-modal entropies, the reader is referred to Ref.~\citenum{glaser24_vib-ent}. As an example, we present the entanglement diagram for one of the excited vibrational states of the formic acid dimer in Fig. \ref{fig:entanglement_fad}.

\begin{figure}[htb]
\begin{lstlisting}[language=Python]
from scine_qcmaquis import QCMaquis

dmrg = QCMaquis()
dmrg.set_entropies() # enables computation of 1 and 2-orbital entropies
dmrg.set_fcidump("LiH_sto3g.fcidump")
norb = 6; nelec = 4; spin = 0
dmrg.run(norb, nelec, spin, fiedler=True)
# 1- and 2-orbital entropies and derived mutual information
s1, s2, mut_inf = dmrg.get_entropies()
\end{lstlisting}
\caption{
Code snippet demonstrating DMRGCI calculation of the LiH diatomic molecule through the Python bindings.
Subsequently, quantum information metrics, namely the 1- and 2-orbital entropies and the mutual information, may be extracted from the optimized MPS
}
\end{figure}

\subsection{Autocorrelation Functions and Population Analysis}
Time-dependent quantities can be extracted along a TD-DMRG propagation using QCMaquis.
This includes the autocorrelation function defined in Eq.~(\ref{eq:autocorr}), from which the spectrum of the system can be derived.

Additionally, for vibronic processes, the population dynamics of the excited states along the PESs can be tracked throughout the time propagation.
The Python interface provides a convenient method for post-processing TD-DMRG results.
The function \texttt{analyze\_results} from Fig.~\ref{fig:analyze} generates plots for the provided measurements.
\begin{figure}[htb]

\begin{lstlisting}[language=Python]
from scine_qcmaquis import QCMaquis

dmrg = QCMaquis(model="vibronic", elec_states=2, vib_modes=4)
# define initial MPS
dmrg.init_mps(
    init_type="coherent",
    init_string="1,0,0,0,0,0|0,1,0,0,0,0", 
    init_coeffs="0.2,0.8"
)
# set integral file
dmrg.set_fcidump("benzoic_acid_vibronic_model.fcidump")
# enable measuring of population dynamics
dmrg.measure_population()
# evolve for 20 femtoseconds
dmrg.evolve(t_step=1, n_steps=20, t_units="fs")
# generate and save figures for the desired measurements
measurements = ["autocorrelation", "spectrum", "population"]
dmrg.analyze_results(measurements)
\end{lstlisting}
\caption{
Code snippet for time evolution and automatic spectrum visualization.
}
\label{fig:analyze}
\end{figure}
The populations and the autocorrelation functions, read from the results file, are plotted and saved in the directory containing the results file.
In addition, the unshifted spectrum under the assumption of a constant dipole moment can be automatically computed from the autocorrelation function.

\section{Technical Aspects}\label{sec:technical}

\subsection{Input and Output}

The standalone version of QCMaquis expects, in addition to the input file, an integral file with all integrals encoding the second-quantized Hamiltonian of the system.
This file is either specified in a text or binary format using the \texttt{integral\_file} or \texttt{integral\_binary} keywords, respectively.
The expected format of the integral file depends on the type of Hamiltonian and is documented in the QCMaquis manual.

In addition to the calculation's output printed to standard output, QCMaquis relies on the HDF5 file format for storing the results of the simulation.
This comprises the \texttt{results\_file}, containing relevant intermediate quantities, such as the bond dimension and energy of each sweep, as well as the requested properties of the final MPS.
The simulation checkpointing mechanism in QCMaquis also relies on the HDF5 file format by storing the MPS at the end of each sweep to disk.
Aggregated into a directory, defined by \texttt{chkp\_file} keyword in the input file, each site tensor is stored in its individual HDF5 file.
This enables users to restart calculations from previous states and perform post-calculation analyses on the properties of the MPS.
The \texttt{results\_file} and \texttt{chk\_file} keywords are optional, and if omitted, the results of the simulation are not stored to a file, and checkpointing will not be performed.

\subsection{Lattice Ordering}
An aspect that can have a significant impact on the performance of DMRG is the mapping of the DoF to the one-dimensional tensor lattice.
In particular, a judicious mapping can significantly reduce the size of the bond dimension required for representing the wavefunction solution~\cite{chan02}.
The default ordering obeys the orbital ordering defined by the FCIDUMP file.
However, this mapping can be overwritten by manually providing a comma-separated string of integers corresponding to the DoF indices in the integral file with the \texttt{orbital\_order} input parameter.
Alternatively, the orbitals may be automatically reordered based on quantum information measures~\cite{Barcza2011Jan} from a partially converged DMRG calculation using the Fiedler ordering.
We recommend enabling this option by default for all electronic structure calculations.

Also, for vibrational models, the ordering of vibrational DoF to the MPS lattice sites is defined by the ordering provided in the integral file.
Typically, for the Watson Hamiltonian, which maps each normal mode to a lattice site, the ordering is chosen in ascending harmonic frequency.
The $n$-mode mapping groups modals relating to a particular mode sequentially, following the same convention of increasing frequency.

Two distinct lattice orderings for vibronic Hamiltonians are available in QCMaquis.
The first is the intertwined sorting, where an electronic site is followed by the vibrational sites associated with that electronic site.
This provides an intuitive ordering for the excitonic Hamiltonian, as it describes a chain of connected monomer units with their own electronic states and vibrational modes grouped together.
The other option places the electronic sites at the beginning of the lattice, followed by the remaining vibrational sites, which is the standard ordering for DMRG calculations with the vibronic Hamiltonian.
These two lattices are depicted in Fig.~\ref{fig:vibronic-lattice}.
The keyword \texttt{vibronic\_sorting} in the input file determines which of the two ordering options is used for the calculations.
The format of the integral file determines the ordering of the individual DoFs for the chosen lattice type.
However, typically, the DoFs are sorted with increasing energy.

In contrast to the electronic structure case, the effects of different lattice orderings for vibrational and vibronic DMRG calculations have not yet been thoroughly investigated.
For vibrational systems, the preliminary steps can be found in our study from Ref.~\cite{glaser24_vib-ent}, which analyzed quantum information metrics extracted from vibrational MPS.
Further studies, including for vibronic systems, will be the subject of future work.

\subsection{MPS Initialization}

QCMaquis can restart calculations from a previous checkpoint by providing by specifying an existing checkpoint directory to the \texttt{chkpfile} keyword.
Otherwise, there are several options for initializing an MPS from scratch.

\subsubsection{Electronic Structure}
The initial guess of the MPS can be either a random MPS of fixed bond dimension or one corresponding to a single determinant, defined by a comma-separated list of DoF occupancies defined in the \texttt{hf\_{occ}} input parameter.
The latter is simply given by an MPS with a bond dimension of one.
To generate the Hartree--Fock determinant, the doubly occupied determinants need to be set to 4, corresponding to doubly occupied, and the unoccupied determinants need to be set to 1, encoding unoccupied determinants.

\subsubsection{Vibrational and Vibronic Structure}
For vibrational and vibronic Hamiltonians, the MPS is usually initialized by a single ONV that describes the particles in the selected vibrational basis state and on the preferred electronic state in the case of vibronic calculations.
The DoFs of this ONV can correspond to occupancies of harmonic oscillator basis functions or vibrational self-consistent field reference modals in the case of vibrational calculations with the Watson or $n$-mode Hamiltonian, respectively.
Furthermore, the MPS can be initialized as a coherent superposition of ONVs.

\section{Automatic Selection of Active Orbital Spaces with AutoCAS}\label{sec:autocas}
Even though the maximum number of orbitals in DMRG can be significantly larger than in CASCI-based methods, it is still too low to capture the entire orbital space, even for small molecular systems. 
In practice, strongly correlated orbitals are selected and form an active space, which is then evaluated by CAS-based methods, to recover static correlation and return a qualitatively correct wavefunction. 
However, a manual orbital choice of the active space is non-trivial and requires deep knowledge of the system. 
Since orbital selection significantly impacts the accuracy of CAS-based methods, numerous strategies have been proposed~\cite{Pulay1988, Bofill1989, Tishchenko2008, Bao2016, Bao2017, Sayfutyarova2017, Bao2018, Khedkar2019, Sayfutyarova2019, Jeong2020, Li2020, Gieseking2021, Golub2021, Khedkar2021, Lei2021, Levine2021, Oakley2021, King2021, Weser2022, Casanova2022, Cheng2022, King2022, Kaufold2023, Golub2023} to guide or automate this process.

A special approach is the autoCAS algorithm~\cite{stein16_automatedselection, Stein2017Apr, Stein2016, stein19_autocas-implementation} developed by our group, which utilizes single-orbital entropies from an unconverged (in terms of bond dimension and the number of sweeps) DMRG wavefunction. 
For the selection of the active space for multi-configurational systems, the relative magnitude of the single-orbital entropies is used to identify strongly correlated orbitals.

Even though the algorithm does not require a converged DMRG wavefunction for the selection of the active space, the number of orbitals is still limited. 
Hence, an initial active space is selected first that consists of all valence orbitals of the system of interest. 
The single orbital entropies~\cite{Legeza2003} of this active space---obtained via the poorly converged DMRG calculation---are then used to select the orbitals for the final production active space. 
For most post-CAS methods, it is crucial to have a balanced active space\cite{Boguslawski2012, Stein2016}, meaning that all strongly correlated orbitals need to be included in the active space. 
To ensure this balancing, the autoCAS algorithm selects the orbitals based on plateaus in the single orbital entropies, such as the ones in Fig.~\ref{fig:threshold_bad}. 
In order to generate these plateaus, the single orbital entropies are first sorted based on their magnitude. 
Since the initial active space is generally larger than the required active space, many orbitals are weakly correlated and have small single orbital entropies, while others, such as the ones responsible for bond-breaking processes, have large orbital entropies. 
This difference in magnitude forms a plateau when the number of orbitals is plotted against the single orbital entropy. 
The number of orbitals that generate this plateau are the ones that are included in the final active space. 
This algorithm ensures that every strongly correlated orbital is included in the active space. 

In cases where the active space is larger than the maximum number of orbitals that can be treated with DMRG, the large CAS protocol\cite{stein19_autocas-implementation} can be applied. 
In this protocol, the initial active space is first divided into the occupied and virtual orbitals. 
Both of these orbital spaces are, then, further divided into a series of subspaces containing only a subset of the respective orbitals. 
Subsequently, every subspace from the occupied orbitals is combined with every subspace from the virtual orbitals to generate sub-active spaces, in which the single-orbital entropies are evaluated.
The final approximate single-orbital entropy for each orbital is chosen as the maximum single-orbital entropy obtained from these sub-active spaces. 
The large CAS protocol can, thus, be utilized to determine active spaces in much larger systems than could be conventionally treated with DMRG, since with this protocol, DMRG calculations only need to be performed in each subspace.

The program SCINE autoCAS\cite{autocas231, Weymuth2024} includes the previously described algorithm, as well as different workflows that can automatically perform active space calculations. 
The package supports multiple electronic structure backends for performing operations such as the orbital optimization and entropy evaluation of the initial and final active spaces.
In addition to the interface to the existing interface with OpenMolcas, a new interface to the PySCF\cite{Sun2020Jul} package has been developed, which utilizes the PySCF interface of QCMaquis to carry out the required DMRG calculation.

\section{Conclusions and Outlook}
The QCMaquis program has evolved into a flexible software for DMRG calculations with extensive functionality for a wide range of applications.
What makes QCMaquis special is its versatility, reaching beyond standard electronic structure calculations to the description of anharmonic vibrational systems, and the resolution of vibronic structures in absorption spectra to TD-DMRG quantum dynamics simulations.
In addition, QCMaquis has pioneered unique techniques unavailable in other packages, including pre-Born–Oppenheimer nuclear-electronic DMRG and transcorrelated DMRG methods.

Major recent advancements, compared to earlier versions of the program, have been the availability of arbitrary basis functions in vDMRG through the $n$-mode Hamiltonian formalism, enhanced insights into correlation patterns in vDMRG through the exposure of modal quantum information measures, and the inclusion of the Frenkel excitonic Hamiltonian.
The application of the new features has been demonstrated in tailored case studies in this work.
The latest release also exposes the functionality of QCMaquis through the newly developed Python bindings, which enable its use as an active space solver for PySCF.

QCMaquis 4.0 marks a milestone in the development of the software. However, future work is already underway and will focus, for instance, on the development of a vibronic model incorporating the $n$-mode Hamiltonian to enable the use of generic basis sets in vibronic time-dependent calculations. This approach offers enhanced flexibility and accuracy for describing complex molecular systems by accounting for anharmonic effects and higher-order couplings.
Another key future goal will be the inclusion of finite-temperature effects in the DMRG algorithm, which would facilitate the exploration of temperature-dependent properties in chemical systems. Moreover, ongoing efforts will aim at extending the software's interfaces to other libraries, including the development of a multi-state extension for the CASPT2 implementation in OpenMolcas.

\begin{acknowledgement}

  We gratefully acknowledge financial support from the Swiss National Science Foundation through Grant No. 200021\_219616.

\end{acknowledgement}

\providecommand{\latin}[1]{#1}
\makeatletter
\providecommand{\doi}
  {\begingroup\let\do\@makeother\dospecials
  \catcode`\{=1 \catcode`\}=2 \doi@aux}
\providecommand{\doi@aux}[1]{\endgroup\texttt{#1}}
\makeatother
\providecommand*\mcitethebibliography{\thebibliography}
\csname @ifundefined\endcsname{endmcitethebibliography}
  {\let\endmcitethebibliography\endthebibliography}{}

\end{document}